\documentclass[cernpreprint, atlasdraft=true, texlive=2016, USenglish, coverpage=false, twocolumn, texmf]{atlasdoc}
 
\usepackage{atlaspackage}
\usepackage{atlasbiblatex}
 
\usepackage{atlasphysics}

\newcommand{\PbPb}{\mbox{Pb+Pb}}
\newcommand{\pbpb}{\mbox{Pb+Pb}}

\newcommand{\Rnmax}{$R_{n}^{\mathrm{max}}$}                      
\newcommand{\nth}{$n^\mathrm{th}$}

\newcommand{\dphin}{$\Delta \phi_n$}
\newcommand{\Psin}{$\Psi_n$}
\newcommand{\Psitwo}{$\Psi_2$}
\newcommand{\Psithree}{$\Psi_3$}
\newcommand{\Psifour}{$\Psi_4$}

\newcommand{\vn}{\mbox{$v_{n}$}}
\newcommand{\vtwo}{\mbox{$v_{\mathrm{2}}$}}
\newcommand{\vthree}{\mbox{$v_{\mathrm{3}}$}}
\newcommand{\vfour}{\mbox{$v_{\mathrm{4}}$}}

\newcommand{\Ncoll}{\mbox{$N_{\mathrm{coll}}$}}

\newcommand{\Njet}{\mbox{$N_{\mathrm{jet}}$}}

\newcommand{\Npart}{\mbox{$N_{\mathrm{part}}$}}

\newcommand{\sqrtsnn}{\mbox{$\sqrt{s_{_\text{NN}}}$}}

\newcommand{\pythiaeight}{\textsc{Pythia}\,8}

\newcommand{\RTwo}{\mbox{$R= 0.2$}}

\newcommand{\TAA}{\mbox{$T_{\mathrm{AA}}$}}

\newcommand{\ETfcal}{\mbox{$\Sigma E_{\mathrm{T}}^{\mathrm{FCal}}$}}

\newcommand{\pTjet}{\mbox{$p_{{\mathrm{T}}}$}}

\newcommand{\pTtrk}{\mbox{$p_{{\mathrm{T}}}^{\mathrm{ch}}$}}

\newcommand{\pTtrue}{\mbox{$p_{\mathrm{T}}^{\mathrm{truth}}$}}

\newcommand{\ETjet}{\mbox{$\ET^{\mathrm{jet}}$}}

\newcommand{\ptjet}{\mbox{$p_{\mathrm{T}}^{\mathrm{jet}}$}}

\graphicspath{{logos/}{figures/}}
 
\usepackage{ANA-HION-2020-09-PAPER-defs}

\clearpage
 
\AtlasTitle{Measurements of azimuthal anisotropies of jet production in Pb+Pb collisions at $\sqrt{s_\mathrm{NN}} =$~5.02~TeV with the ATLAS detector}
 
\AtlasAbstract{
The azimuthal variation of jet yields in heavy-ion collisions provides information about the path-length dependence
of the energy loss experienced by partons passing through the hot, dense nuclear matter known as the quark--gluon plasma.
This paper presents the azimuthal anisotropy coefficients $v_2$, $v_3$, and $v_4$ measured for jets in Pb+Pb collisions
at $\sqrt{s_\mathrm{NN}} =$~5.02~TeV using the ATLAS detector at the LHC.
The measurement uses data collected in 2015 and 2018, corresponding to an integrated luminosity of 2.2~nb$^{-1}$.
The $v_n$ values are measured as a function of the transverse momentum of the jets between 71~GeV and 398~GeV and the event centrality. A nonzero value of $v_2$ is observed
in all but the most central collisions. The value of $v_2$ is largest for jets with lower transverse momentum, with values up to 0.05 in mid-central collisions.
A smaller, nonzero value of $v_3$ of approximately 0.01 is measured with no significant dependence on jet $p_\mathrm{T}$ or centrality, suggesting that
fluctuations in the initial state play a small but distinct role in jet energy loss. No significant deviation of $v_4$ from zero is observed in the measured kinematic region.
}
 
\author{The ATLAS Collaboration}
 
\AtlasRefCode{HION-2020-09}
\AtlasJournal{PRC}
\AtlasJournalRef{Phys. Rev. C 105 (2022) 064903}
\AtlasDOI{10.1103/PhysRevC.105.064903}
 
\PreprintIdNumber{CERN-EP-2021-218}
\clearpage
\hypersetup{pdftitle={ATLAS document},pdfauthor={The ATLAS Collaboration}}
 
\begin{document}
 
\maketitle

\clearpage
\section{Introduction}
The primary physics aim of the heavy-ion program at the Large Hadron Collider (LHC) is to produce
and study the quark--gluon plasma (QGP), the high-temperature state of quantum-chromodynamic matter
in which quarks and gluons are no longer confined within protons and neutrons (for a recent review, see
Ref.~\cite{Busza:2018rrf}).
Measurements of jets produced in the early stages of
heavy-ion collisions provide information about the short-distance-scale
interactions of high-energy partons with the QGP.  The overall rate of jets in central Pb+Pb collisions at a given transverse momentum
\pt is found to be about a factor of two lower than expectations based on
\pp\ collisions, up to a \pt of approximately 1~\TeV~\cite{Aaboud:2018twu,CMS:2021vui}.
This suppression can be explained by the downward slope of the jet \pt\ spectrum and the reduction in parton \pt due to energy loss while traversing the QGP.
The energy loss from partons is expected to depend on the length of the QGP region that
the parton traverses.  The geometry of the overlapping nuclei in mid-central collisions
leads to shorter average path lengths if the jet is oriented along the direction of the collision impact
parameter vector\footnote{The impact parameter is the distance between the centers of the colliding nuclei in the plane transverse to the collision axis.}
than if the jet is oriented in the perpendicular direction.  This should
lead to a dependence of the jet yield on the  azimuthal angle~\cite{Wang:2000fq,Gyulassy:2000gk,Shuryak:2001me}.

One key observable in understanding the path-length dependence of energy loss is the azimuthal anisotropy of jets.
The azimuthal distribution of jets is described via a Fourier expansion:
\begin{equation}\nonumber
\frac{{d}N_\mathrm{jet}}{{d}\phi} \propto 1 + 2 \sum_{n=1}^\infty \vn \cos(n(\phi - \Psi_n)),
\end{equation}
where the \vn\ and \Psin\ are the magnitude and orientation of the \nth-order anisotropy, and $\phi$ is
the azimuthal angle of jets.
The \Psin, or event-plane angles, are oriented such that a jet produced in-plane, or along the direction of the event-plane angle, will traverse on average less
QGP than a jet produced out-of-plane, or perpendicular to the event-plane angle.
Similar Fourier expansions are often used to describe the azimuthal variation of the yield of soft particles, which is typically associated with hydrodynamic flow
(see Ref.~\cite{Heinz:2013th}).
It is important to note that at high \pt, hydrodynamic flow is not expected to be the source of azimuthal variation.
Measurements of the \vn\ for high-\pt\ particles have been performed at the Relativistic Heavy Ion Collider (RHIC) \cite{PHENIX:2013yhu,Adams_2005}.
The first measurement of the \vtwo\ for fully reconstructed jets was reported in Ref.~\cite{Aad:2013sla} for
\pbpb\ collisions at $\sqrtsnn = 2.76$~\TeV.
The measured \vtwo\ values were found to be positive for jets with transverse momentum
45--160~\GeV. The \vtwo\ values were found to be smaller in the most central and most peripheral collisions.
This is expected because the second-order eccentricity of the initial state is small in the most central collisions, while in the
most peripheral collisions there is little energy loss in any direction.
A measurement by ALICE using jets reconstructed from charged particles obtained similar
results~\cite{Adam_2016}.
Related measurements by CMS and ATLAS have been performed
with charged particles at high \pt\ in 5.02~\TeV\ \pbpb\ collisions~\cite{Sirunyan:2017pan,HION-2016-06}.  Reference~\cite{Sirunyan:2017pan}
reported positive \vtwo\ values for charged particles with \pt\ up to 60--80~\GeV.
Until now, there have been no measurements of jet \vtwo\
in $\sqrtsnn = 5.02$~\TeV\ \pbpb\ collisions and no measurements
of the higher-order anisotropies, such as \vthree\ and \vfour, of jets in any collision system.
Such measurements could provide new information about how the energy loss depends on path length and the initial collision geometry.
 
Recent calculations have shown that
realistic modeling of both the jet energy loss and the soft fluctuations are necessary to
reproduce the experimental measurements of high-\pt\ particles~\cite{Noronha-Hostler:2016eow}.
Therefore, it is of interest to study observables that are sensitive to both
path-length dependence of the energy loss and fluctuations of the initial collision geometry,
such as the dependence of the jet yield on higher-order eccentricities of the initial state~\cite{Betz:2016ayq}.
 
The results presented here extend the measurement of jet azimuthal anisotropy to higher jet \pt\
than in previous measurements and to a collision energy of $\sqrtsnn = 5.02$~\TeV.
Additionally, higher-order harmonics, \vthree\ and \vfour, are measured.
The measurement utilizes 2.2~nb$^{-1}$ of \pbpb\ data collected at $\sqrtsnn = 5.02$~\TeV\
in 2015 and 2018.
Jets are reconstructed using the \antikt~\cite{Cacciari:2008gp} algorithm with \RTwo.
Compared to larger-radius jets, these small-radius jets provide improved angular resolution
for the estimation of the jet axis; this improvement is due to the smaller underlying event within the jet cone,
which helps in measuring the angular anisotropies.
The jets used in this analysis are restricted to rapidities $|y|<1.2$.\footnote{ATLAS uses a right-handed
coordinate system with its origin at the nominal interaction point (IP) in the center of the detector, and the $z$-axis along the beam pipe.
The $x$-axis points from the IP to the center of the LHC ring, and the $y$-axis points upward.
Cylindrical coordinates $(r,\phi)$ are used in the transverse plane, $\phi$ being the azimuthal angle around the $z$-axis.
The pseudorapidity is defined in terms of the polar angle $\theta$ as $\eta=-\ln\tan(\theta/2)$.
The rapidity is defined as $y = 0.5\ln[(E + p_z)/(E-p_z)]$ where $E$ and $p_z$ are the energy and $z$-component of the momentum along the beam direction, respectively.
Transverse momentum and transverse energy are defined as $\pt = p \sin\theta$ and $\ET = E \sin\theta$, respectively.
The angular distance between two objects with relative differences $\Delta \eta$ in pseudorapidity and $\Delta \phi$ in azimuth is given by $\Delta R = \sqrt{(\Delta \eta )^2 + (\Delta \phi)^2}$.}
The observed event-plane angles, $\Psi_n^\mathrm{obs}$, are reconstructed using the transverse energy measured over
$4.0<|\eta|<4.9$ as described in Section~\ref{sec:analysis}
and the $v_n^\mathrm{obs}$ values are extracted by fitting independently for each order, $n$:
\begin{equation}\nonumber
\frac{{d}N_{\mathrm{jet}}(\pt,\Delta \phi_n)}{{d}\Delta \phi_n} \propto 1 + 2 v_n^\mathrm{obs} \cos(n \Delta \phi_n ).
\end{equation}
Here $N_{\mathrm{jet}}(\pt,\Delta \phi_n)$ represents the number of jets for a given \pt\ and $\Delta \phi_n$ selection, where $\Delta \phi_n$ is defined as $\Delta \phi_n = |\Psi_n^\mathrm{obs} - \phi|$.
\section{ATLAS detector and trigger}
The measurement presented in this paper is performed using the ATLAS calorimeter, inner detector, trigger, and data acquisition
systems~\cite{PERF-2007-01}. An extensive software suite~\cite{ATL-SOFT-PUB-2021-001} is used in the reconstruction and analysis of real and simulated data, in detector operations,
and in the trigger and data acquisition systems of the experiment.
 
The calorimeter system consists of a sampling liquid-argon (LAr) electromagnetic (EM) calorimeter covering
$|\eta|<3.2$, a steel--scintillator sampling hadronic calorimeter covering $|\eta| <1.7$, LAr hadronic calorimeters covering $1.5 < |\eta| < 3.2$,
and two LAr forward calorimeters~(FCal) covering $3.2 < |\eta| < 4.9$. The EM calorimeters are segmented longitudinally in shower depth into
three layers with an additional presampler layer covering $|\eta| < 1.8$. The hadronic calorimeters have three sampling layers longitudinal in shower
depth in $|\eta| < 1.7$ and four
sampling layers in $1.5 < |\eta| < 3.2$, with a slight overlap in $\eta$.
 
The inner detector measures charged particles within the pseudorapidity interval $|\eta|<2.5$ using a combination of silicon pixel detectors,
silicon microstrip detectors (SCT), and a straw-tube transition radiation tracker (TRT), all immersed in a 2~T axial magnetic field~\cite{PERF-2007-01}.
Each of the three detectors is composed of a barrel and two symmetric endcap sections. The pixel detector is composed of four layers including
the insertable B-layer~\cite{ATLAS-TDR-19,PIX-2018-001}. The SCT barrel section contains four layers of modules with sensors on both sides,
and each endcap consists of nine layers of double-sided modules with radial strips. The TRT contains layers of staggered straws interleaved with
the transition radiation material.
 
The zero-degree calorimeters (ZDCs) are located symmetrically at $z = \pm140$~m and cover $|\eta| > 8.3$.
The ZDCs use tungsten plates as absorbers, and quartz rods sandwiched between the tungsten plates as the active medium.
In \PbPb\ collisions, the ZDCs primarily measure ``spectator'' neutrons that do not interact hadronically when the incident nuclei collide.
A ZDC coincidence trigger is implemented by requiring the pulse height from both ZDCs to be above a threshold which is set to accept
the signal corresponding to the energy deposition from a single neutron.
 
ATLAS uses a two-level trigger system. The first-level trigger is hardware-based and implemented with custom electronics.
It is followed by the software-based high-level trigger (HLT)~\cite{TRIG-2019-04}.

\section{Data and event selection}
This analysis uses data from \pbpb\ runs at $\sqrtsnn = 5.02$~\TeV\ collected by the ATLAS detector in 2015 and 2018. Events were selected online
by a combination of jet triggers. In the HLT, they require a jet with radius parameter $R = 0.4$ with \pt\ greater than 50, 60, 75, 85, or 100~\GeV.
The 100~\GeV\ jet trigger sampled the full integrated luminosity of 0.5~nb$^{-1}$ in 2015 and 1.7~nb$^{-1}$ in 2018, while the lower-threshold triggers
were prescaled. The data are selected by using each trigger in the region of jet \pt\
for which the HLT triggers are more than 99\% efficient, where the efficiency is calculated using reconstructed $R = 0.2$ jets.
To populate regions with lower jet \pt, events passing minimum-bias (MB) triggers are also included.
More details about the jet triggering in heavy-ion collisions can be found
in Ref.~\cite{ATL-DAQ-PUB-2019-001}.
 
The offline event selection requires that
events pass both an in-time pileup cut based on the ZDC energy and the total transverse energy in the FCal, and an out-of-time pileup cut based
on the number of reconstructed tracks in an event and the total transverse energy in the calorimeter.
Here, in-time pileup refers to events with multiple interactions in the same bunch crossing, and out-of-time pileup refers to
events in which energy from a previous bunch crossing affects the energy measured in the calorimeter.
The pileup rejection cuts remove less than 0.5\% of events.
Jets selected offline have $|y| < 1.2$
and \pt\ in the range of 63--501~\GeV, with jets in the ranges 63--71~\GeV\ and 398--501~\GeV\ used to populate, respectively, the underflow and overflow bins in an unfolding procedure to correct for jet energy scale and resolution effects.
The centrality of an event is determined by the sum of transverse energy in the forward calorimeters, \ETfcal.
Centrality percentiles are determined by separating the MB events into percentiles based on the \ETfcal\ in each event, ranging from the most central (smallest impact parameter, highest \ETfcal) to the most peripheral (largest impact parameter, lowest \ETfcal),
as described in Ref.~\cite{HION-2016-04}.
Events are selected with centralities of 0--5\%, 5--10\%, 10--20\%, 20--40\%, and 40--60\%.
 
This analysis uses Monte Carlo (MC) simulations to evaluate the performance of the detector and analysis procedure, and to correct the measured
distributions for detector effects. The detector response in all
MC samples was simulated using \textsc{Geant4}~\cite{Agostinelli:2002hh,Aad:2010ah}. The \PbPb\ MC sample makes use of $7\times10^7$
dijet events from 5.02~\TeV\ $pp$ collisions simulated by \pythiaeight~\cite{Sjostrand:2014zea} with the A14 set of tuned parameters~\cite{ATLAS2014021} and the
\textsc{NNPDF2.3lo} parton distribution functions~\cite{Ball:2012cx}.
Events from the \pythiaeight\ dijet sample are overlayed with events from a dedicated sample of \PbPb\ data events.
This sample was recorded with a combination of the MB trigger and triggers requiring a total energy above 1.5~\TeV\ or 6.5~\TeV\ to enhance the
number of central collisions. The overlay procedure combines the \pythiaeight\ and data events during the digitization step of simulation.
This MC overlay sample was reweighted
on an event-by-event basis such that it has the same \ETfcal\ distribution as the jet-triggered data sample to better represent the centrality distribution
of the data used in this analysis.

\section{Analysis procedure}\label{sec:analysis}
This analysis uses the event-plane method to determine \vn\ coefficients as described in Ref.~\cite{PhysRevC.58.1671} and used in previous measurements~\cite{HION-2017-05, HION-2019-11}.
The geometry of the initial collision can be characterized by a series of observed event-plane angles, $\Psi_n^\mathrm{obs}$,\footnote{The observed event-plane angles are defined as
$\Psi_n = \frac{1}{n} \tan^{-1}\frac{\sum_j E_{\mathrm{T},j}\sin(n\phi_j)}{\sum_j E_{\mathrm{T},j}\cos(n\phi_j)}$ where $E_{\mathrm{T},j}$ is the transverse
energy measured in calorimeter tower $j$
of the forward calorimeters.} determined by the azimuthal variation of transverse energy
in the forward calorimeters. Only the range $|\eta| > 4.0$ of the forward calorimeters is used in this analysis to reduce any bias of the event-plane determination from
jets in the FCal. The resolution of the event-plane angles, $\mathrm{Res}\{\Psi_n\}$, is determined by comparing in each event the values
calculated in the forward and backward sides of the detector as detailed in Ref.~\cite{HION-2016-06}.
The resolution is determined for each bin in centrality, and ranges from approximately 0.6--0.9 for \Psitwo, 0.3--0.6 for \Psithree, and 0.2--0.3 for \Psifour.
 
The jet reconstruction procedures follow those used by ATLAS for previous jet measurements in \PbPb\ collisions~\cite{Aaboud:2018twu,Aad:2013sla}.
Jets are reconstructed using
the \antikt\ algorithm~\cite{Cacciari:2008gp} implemented in the \verb=FastJet= software package~\cite{Cacciari:2011ma}. Jets with
$R=0.2$ are formed by
clustering calorimetric towers of spatial size $\Delta\eta \times\Delta\phi=0.1\times\pi/32$. The energies in the towers are obtained
by summing the energies of calorimeter cells at the electromagnetic energy scale~\cite{PERF-2011-03} within the tower boundaries. A background
subtraction procedure is applied to estimate within each event the underlying event (UE) average transverse energy density,
$\rho(\eta,\phi)$, where the $\phi$ dependence is due to global azimuthal correlations in the particle production from
hydrodynamic flow~\cite{HION-2016-06}. The modulation accounts for the contribution to the UE of the second-, third-, and fourth-order azimuthal anisotropy harmonics
characterized by values of flow coefficients $v_n^\mathrm{UE}$~\cite{HION-2016-06}. Any potential residual effect of the azimuthal variation of the underlying
event on the jet reconstruction is accounted for by the systematic uncertainties described in Section~\ref{sec:systematics}.
The UE is also corrected for $\eta$- and $\phi$-dependent nonuniformities of the detector response
by correction factors derived in MB \PbPb\ data.
 
An iterative procedure is used to remove the impact of jets on the estimated $\rho$ and $v_n^\mathrm{UE}$
values.  The first estimate of the average transverse energy density of the UE, $\rho(\eta)$, is evaluated in 0.1 intervals of $\eta$, excluding those
which overlap with ``seed'' jets. In the first subtraction step, the seeds are defined to be a union of $R = 0.2$ jets and $R=0.4$ track-jets. Track-jets are
reconstructed by applying the \antikt\ algorithm with $R = 0.4$ to charged particles with $\pT > 4$~\GeV. The $R=0.2$ jets must pass a cut on the
value of the tower energy, while the track-jets are required to have $\pT > 7$~\GeV. The background is then subtracted from each tower
constituent and jet kinematics are recalculated. After the first iteration, the $\rho$ and $v_n$ values are updated by excluding from the UE determination
the regions within $\Delta{R} = 0.4$ of both the track-jets and the newly reconstructed $R=0.2$ jets with $\pt > 25$~\GeV. The updated
$\rho$ and $v_n^\mathrm{UE}$ values are used to update the jet kinematic properties in the second iteration.
 
Jet $\eta$- and $\pT$-dependent correction factors derived in simulations are applied to the measured jet
energy to correct for the calorimeter energy response~\cite{Aad:2014bia}. An additional correction based on \textit{in situ} studies of jets recoiling
against photons, \Zboson\ bosons, and jets in other regions of the calorimeter is applied~\cite{HION-2018-06}. This calibration is followed by a
``cross-calibration'' which relates the jet energy scale (JES) of jets reconstructed by the procedure outlined in this section to the JES
in 13~\TeV\ \pp\ collisions~\cite{ATLAS:2015twa}.

So-called ``truth jets'' are defined in the MC sample before detector simulation by applying the \antikt\
algorithm with $R = 0.2$ to stable particles with a proper lifetime greater than 30~ps, but excluding muons and neutrinos, which do not
leave significant energy deposits in the calorimeter.
 
\begin{figure*}[t]
\centering
\subfloat{
\centering
\includegraphics[width=0.5\textwidth]{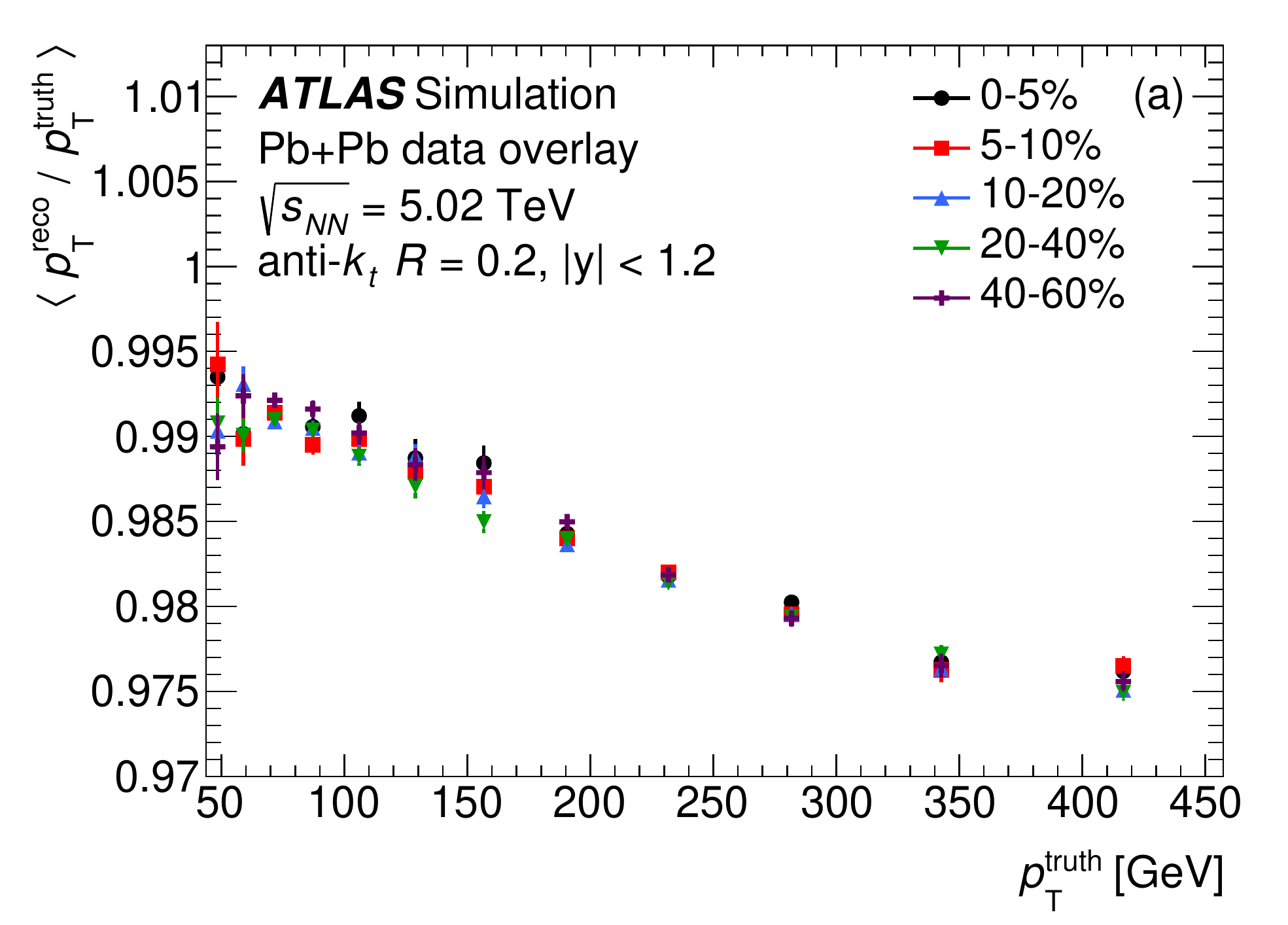}
}
\subfloat{
\centering
\includegraphics[width=0.5\textwidth]{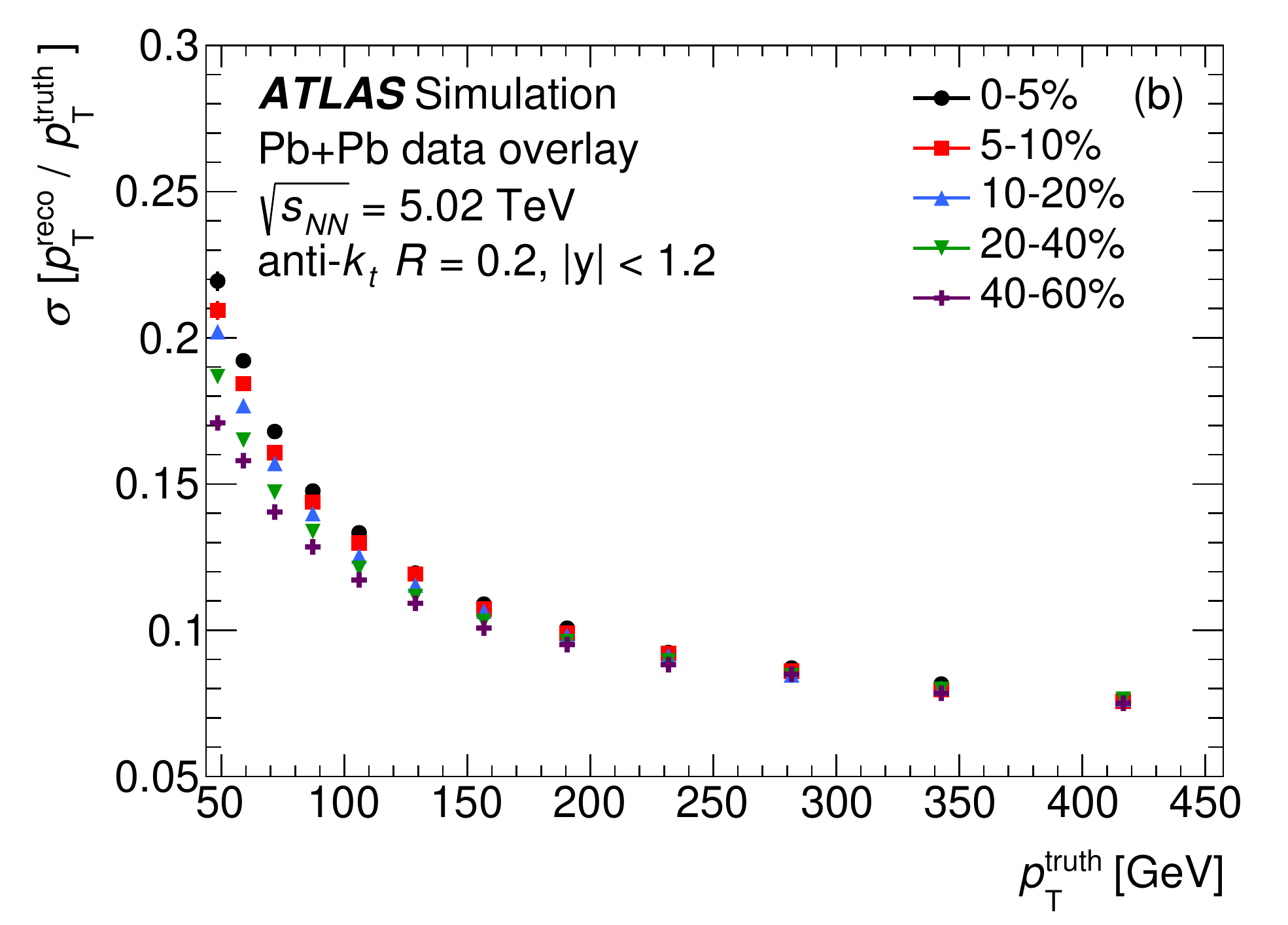}
}
\caption{\label{fig:jesjerpt} The (a) JES and (b) JER
for $R=0.2$ jets in \pbpb\ collisions as a function of $p_\mathrm{T}^\mathrm{truth}$ for centrality selections of 0--5\%, 5--10\%, 10--20\%, 20--40\% and 40--60\%.}
\end{figure*}

\begin{figure*}[t]
\centering
\subfloat{
\centering
\includegraphics[width=0.5\textwidth]{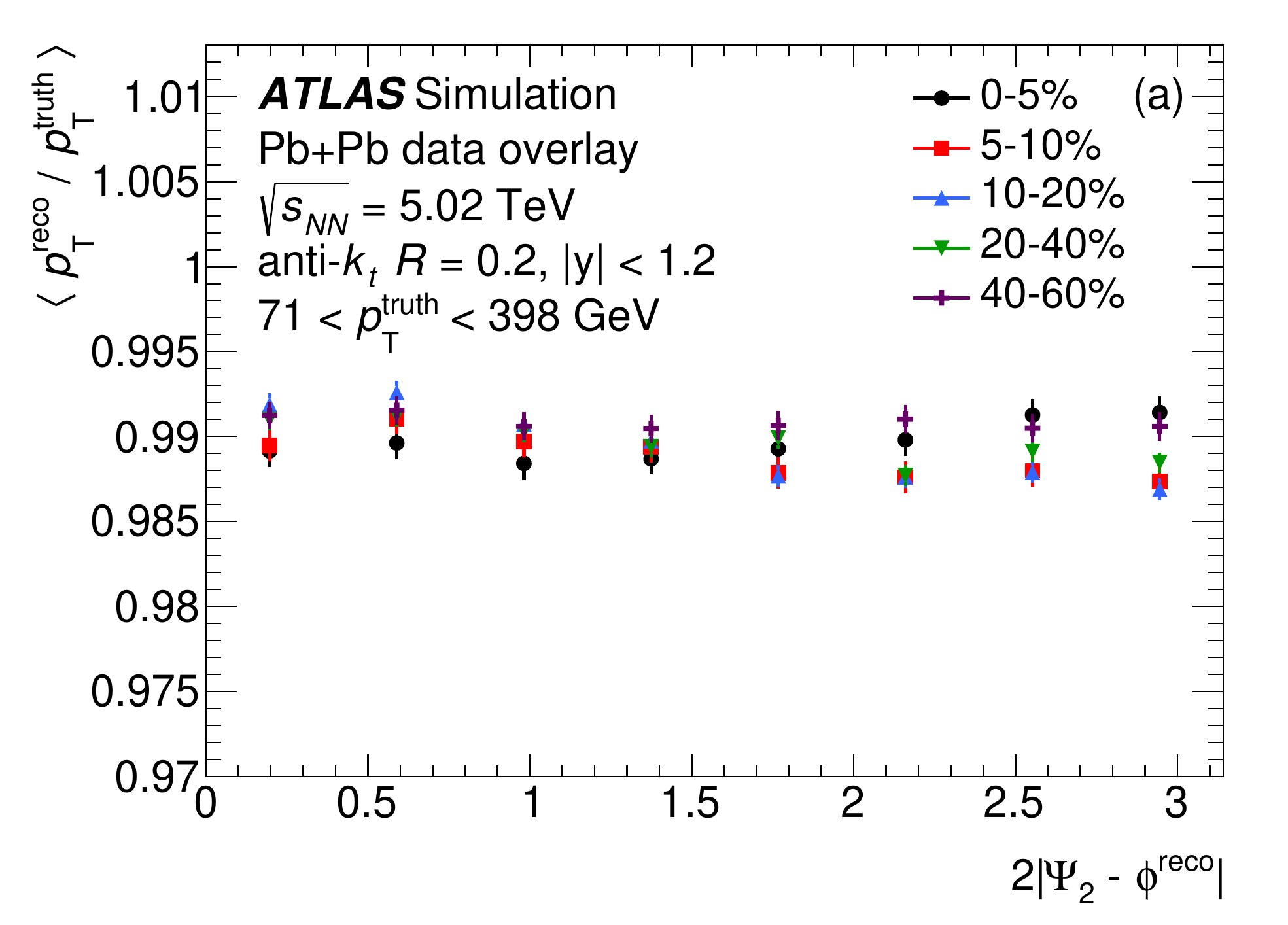}
}
\subfloat{
\centering
\includegraphics[width=0.5\textwidth]{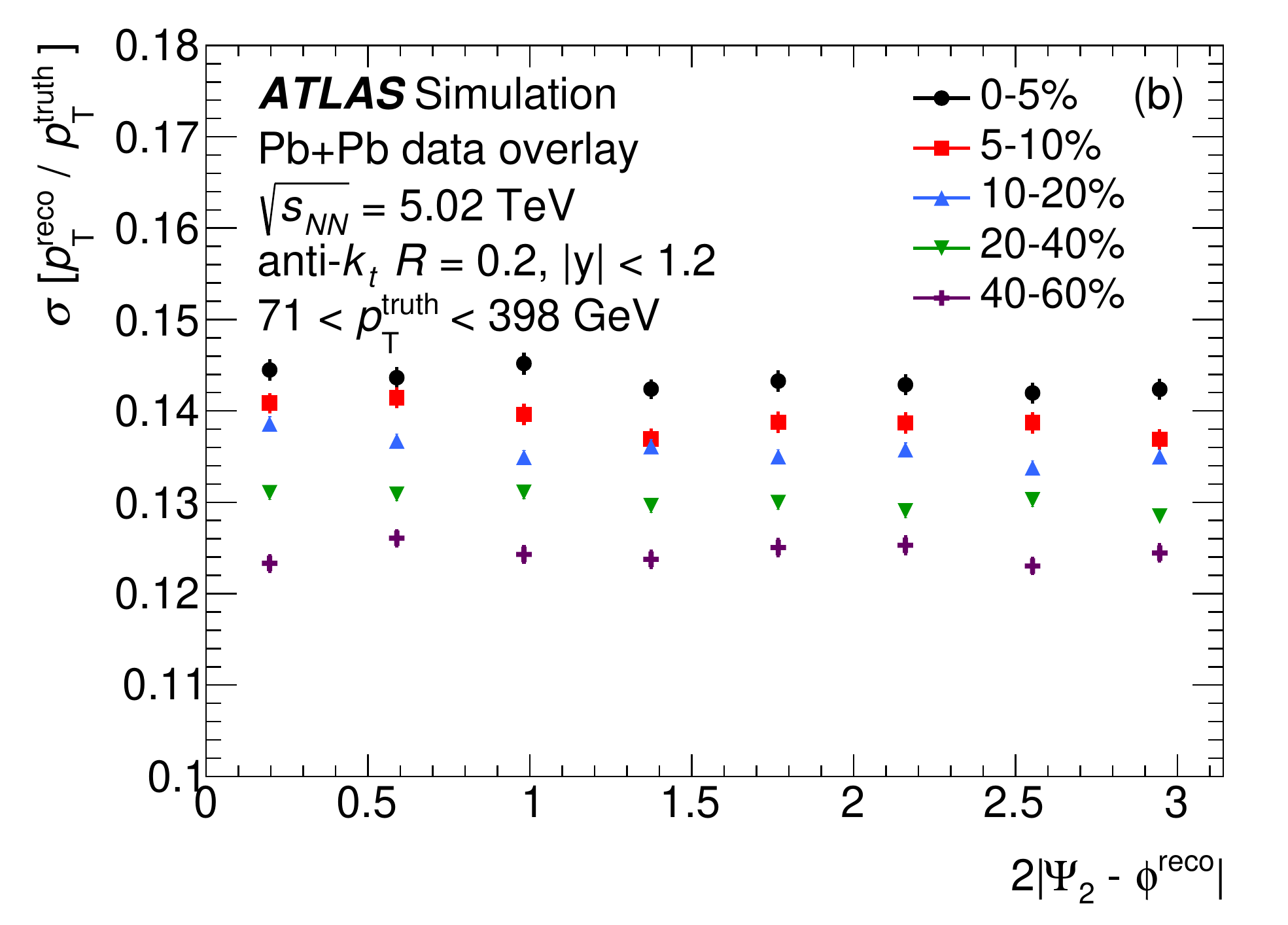}
}
\caption{\label{fig:jesjerdphi} The (a) JES and (b) JER
for $R=0.2$ jets in \pbpb\ collisions as a function of $2|\Psi_2^\mathrm{obs}-\phi^\mathrm{reco}|$ for centrality selections of 0--5\%, 5--10\%, 10--20\%, 20--40\% and 40--60\%.}
\end{figure*}

The JES and jet energy resolution (JER) for \RTwo\ jets are shown in Figure~\ref{fig:jesjerpt} as a function of $p_\mathrm{T}^\mathrm{truth}$.
They are derived by matching each truth jet to the closest reconstructed and calibrated jet from the MC overlay sample within an angular distance of $\Delta R = 0.15$.
The JES and JER are taken to be the means and standard deviations of the $p_\mathrm{T}^\mathrm{reco}/p_\mathrm{T}^\mathrm{truth}$ distributions, respectively.
The JES differs from unity by approximately 1\% at 70~\GeV\ and 2.5\% at 400~\GeV; this deviation
is due to isolation cuts used in the determination of the jet calibration and is corrected for by the unfolding procedure
described below.  The
JES has no significant centrality dependence.  The JER improves with increasing \pt\
and from central to peripheral collisions.
Figure~\ref{fig:jesjerdphi} shows the JES and JER for \RTwo\ jets as a function of the angle between the jet and the observed second-order event-plane angle.
The dependence of the JES on this angle is smaller than its dependence on \pt,
with variations up to approximately 0.5\% between in-plane and out-of-plane
jets. The rapidity range used in this measurement, $|y| < 1.2$, is selected to minimize the JES dependence on the angle with respect to the event plane.
The JER also shows a small dependence on the angle between the jet and the second-order event-plane angle,
with the resolution of in-plane jets up to 0.5\% larger than that for out-of-plane jets.
 
The jet yield is determined as a
function of \pt, centrality, and \dphin. For each centrality and \dphin\ selection, the jet \pt\ spectra are unfolded to correct for jet energy scale
and resolution effects using a one-dimensional
Bayesian unfolding~\cite{DAgostini:1994fjx} as implemented in the \verb=RooUnfold= package~\cite{Adye:2011gm}.
The response matrices are filled using spatially matched truth jet and reconstructed jet pairs from the MC overlay sample.
The response matrices are reweighted in truth \pt\ by the ratio of the \pt\ spectra in data to that in the reconstructed
MC sample, such that the \pt\ spectra in the response matrices better represent those in the data.
The reweighting is done separately in each \dphin\ bin, such that the response matrices include the same
modulation as seen in the raw data.
The unfolding is performed using three iterations, which was found to minimize the combination of the statistical uncertainty and relative
bin migration for subsequent iterations. The data are not unfolded to correct for the angular resolution of the jets, which is found to be
small compared to the size of the \dphin\ binning.
 
For each selection in \pt, centrality and harmonic value $n$, a function is fitted to the
unfolded \dphin\ distributions to extract the $v_n^\mathrm{obs}$ values. The fit function is:
\begin{equation}
A(1+2 v_{n}^\mathrm{obs} \cos (n\Delta \phi_n))  \nonumber
\end{equation}
where the overall normalization $A$ and the value of $v_n^\mathrm{obs}$ are the free parameters in the fitting procedure.
The fitted  $v_n^\mathrm{obs}$ values are then corrected for the finite event-plane resolution as described in Ref.~\cite{PhysRevC.58.1671}, where $v_n =  v_n^\mathrm{obs}/\mathrm{Res}\{\Psi_n\}$.
In addition to the \vn\ measurements differential in jet \pt, the values are also obtained in an inclusive \pt\ bin for jets with $71 < \pt < 398$~\GeV,
following the same procedure as used in the differential measurement.
\section{Systematic uncertainties}\label{sec:systematics}
The systematic uncertainties in this measurement arise from the JES and JER, the
unfolding procedure,
and the biasing of the event plane by a forward-produced jet correlated with the jet of interest.
The systematic uncertainties presented in this section are given in terms of the absolute change to the measured \vn\ values.
For each uncertainty component the entire analysis procedure
is repeated with the variation under consideration and the uncertainty
contributions are added in quadrature to obtain the total systematic uncertainty in the measurement.
 
The systematic uncertainty in the JES has six parts. First, a centrality-independent baseline component is determined from \textit{in situ}
studies of the calorimeter response to jets reconstructed with the procedure used in 13~\TeV\ \pp\ collisions~\cite{PERF-2011-03,Aaboud:2017jcu}.
A second, centrality-independent component accounts for the relative energy scale difference between the jet reconstruction procedures
used in this analysis and those in 13~\TeV\ \pp\ collisions~\cite{ATLAS:2015twa}. Potential inaccuracies in the MC sample in the description
of the relative abundances of jets initiated by quarks and gluons and of the calorimetric response to quark and gluon jets
are accounted for by the third component. The fourth, centrality-dependent, component accounts for modifications of the parton shower due to quenching and thus
possibly a different detector response to jets in \PbPb\ collisions that is not modeled by the MC simulation. It is evaluated by the method used
for 2015 and 2011 data~\cite{ATLAS:2015twa}, which compares the jet \pt\ measured in the calorimeter and the sum of the transverse momenta of charged particles within the jet, in both the data and MC samples. The charged particles are selected with $\pt > 4$~\GeV\ to remove effects of the UE.
The size of the centrality-dependent uncertainty in the JES reaches 1.2\% in the most central collisions.
An additional, centrality-independent component of 0.5\% is included to account for potential year-to-year differences observed between the peripheral \pbpb\ data
taken in 2018 and the \pp\ collision data taken in 2017 which is used for the calibration.
The systematic uncertainties from the JES discussed above are derived for $R=0.4$ jets. The fifth component does not depend on collision centrality and
it accounts for the potential difference in
uncertainties between  $R = 0.4$ and $R = 0.2$ jets. This uncertainty is assessed by comparing the ratio of \pt\ for matched $R = 0.2$
and $R = 0.4$ jets measured in data and the MC sample. The size of this JES uncertainty is approximately 1\%.
Each component is varied separately by $\pm1$ standard deviation in MC samples, applied
as a function of \pT\ and $\eta$, and the response matrices are recomputed. The data are then unfolded with the modified matrices.
Because the measurement is sensitive only to the relative variation in yields as a function of \dphin, the measured \vn\ values are insensitive
to these JES uncertainties that do not depend on \dphin\ and therefore these are subdominant uncertainties.
 
The sixth uncertainty in the JES comes from a potential variation of the scale as a function of the angle between the jet and the event plane.
The maximum size of the variations is determined by comparing the jet \pt\ measured in the calorimeter and the
sum of the transverse momenta of charged particles within the jet, as a function of \dphin, in both the data and MC samples.
The \vn\ due to potential variations in the JES, $v_n^{\mathrm{JES}}$, is determined by modifying the jets in the MC sample for
different values of \dphin\ using the comparison of the calorimeter and track measurements and measuring the resulting \vn.
The data in each \dphin\ bin are then scaled by $1+2v_n^{\mathrm{JES}}\cos(n \Delta\phi_n)$ and
fit to extract the systematic variation. Because this measurement is only sensitive to the relative jet yields as a function of \dphin\ and not the overall
scale of the yields, systematic variations that vary as a function of \dphin\ will result in a larger uncertainty in the \vn\ than variations which only
depend on the \pt\ of a jet, such as those described above. Therefore, the uncertainty in the variation of the scale as a function of the angle between
the jet and the event plane is the dominant uncertainty in the JES for this measurement. In 20--40\% central collisions for jets with $71 < \pt < 79$~GeV
where the measured \vtwo\ is largest, this uncertainty accounts for approximately 95\% of the total uncertainty on the \vtwo\ due to the JES. For jets in
the same centrality collisions with $316 < \pt < 398$~GeV, this uncertainty accounts for approximately 80\% of the total uncertainty on the \vtwo\ due to the JES.
In 0--5\%, 5--10\%, and 10--20\% central collisions, this uncertainty accounts for >80\% of the total uncertainty on the \vthree\ and \vfour\ due to the JES for
the full kinematic range of the measurement.
 
The uncertainty due to the JER is evaluated by repeating the unfolding procedure with modified response matrices, where an additional contribution is added to the
resolution of the reconstructed \pT\ in the MC sample using a Gaussian smearing procedure.
The smearing factor is evaluated using an \textit{in situ} technique in 13~\TeV\ \pp\ data that involves studies of dijet
energy balance~\cite{Aad:2012ag,PERF-2014-02}.
Further, an uncertainty is included to account for differences between the tower-based jet reconstruction and the jet reconstruction used in
analyses of 13~\TeV\ \pp\ data, as well as differences in calibration procedures.
Similarly to the JES, an additional uncertainty is assigned to the JER to account for differences between $R = 0.2$  and $R = 0.4$ jets.
The resulting uncertainty from the JER is symmetrized.
 
The final uncertainty in the JER comes from a potential variation of the resolution as a function of the angle between the jet and the event plane
due to the increased size of the UE in-plane compared to out-of-plane.
The size of the UE is correlated with the size of the fluctuations of the UE which can lead to too small or too large a subtraction and increase the JER.
The \vn\ due to potential variations in the JER, $v_n^{\mathrm{JER}}$, is determined by adding an additional contribution to the JER of the jets in the MC sample for
different values of \dphin\ and	measuring the resulting	\vn.
This additional contribution to the JER is determined by correlating the fluctuations in the UE with the size of the UE in data.
The unfolded data in each \dphin\ bin are then scaled by $1+2v_n^{\mathrm{JER}}\cos(n \Delta\phi_n)$ and
fit to extract the systematic variation. The variations in the JER have a minimal effect on the measured \vn\ values.
 
The uncertainty in the unfolding procedure was determined by unfolding the data with response matrices that had not been reweighted to match the \pt\ spectra in data
as described in Section~\ref{sec:analysis} and fitting the unfolded data to obtain new \vn\ results.
The deviation from the nominal unfolding result was symmetrized and taken as the systematic uncertainty contribution.
 
The uncertainty in the event-plane resolution as determined in Ref.~\cite{HION-2019-11} was found to be negligible in comparison with other
uncertainties and is not included.
However, it is possible for a jet correlated with the jet of interest to bias the event plane
if some of its energy is in the FCal.  An estimate of the size of this effect
was determined from the MC samples.
The MC samples were produced without a correlation between the dijets in \pythiaeight\
and the \Psin\ angles in the overlaid data event.  Therefore, the measured \vn\ of jets coming from the \pythiaeight\ event
should be zero, and any nonzero \vn\ values are caused by some events
having their event-plane determination biased by a jet from the MC sample.
The size of the effect that jets biasing the event-plane angles have on the \vn\ measured in data is estimated using the
\vn\ values found in the MC sample. The azimuthal modulation of the jet yields in the MC sample is subtracted from that in the unfolded data
and the resulting \vn\ values are taken as the systematic variations.
 
\begin{figure*}[ht]
\centering
\subfloat{
\centering
\includegraphics[width=0.5\textwidth]{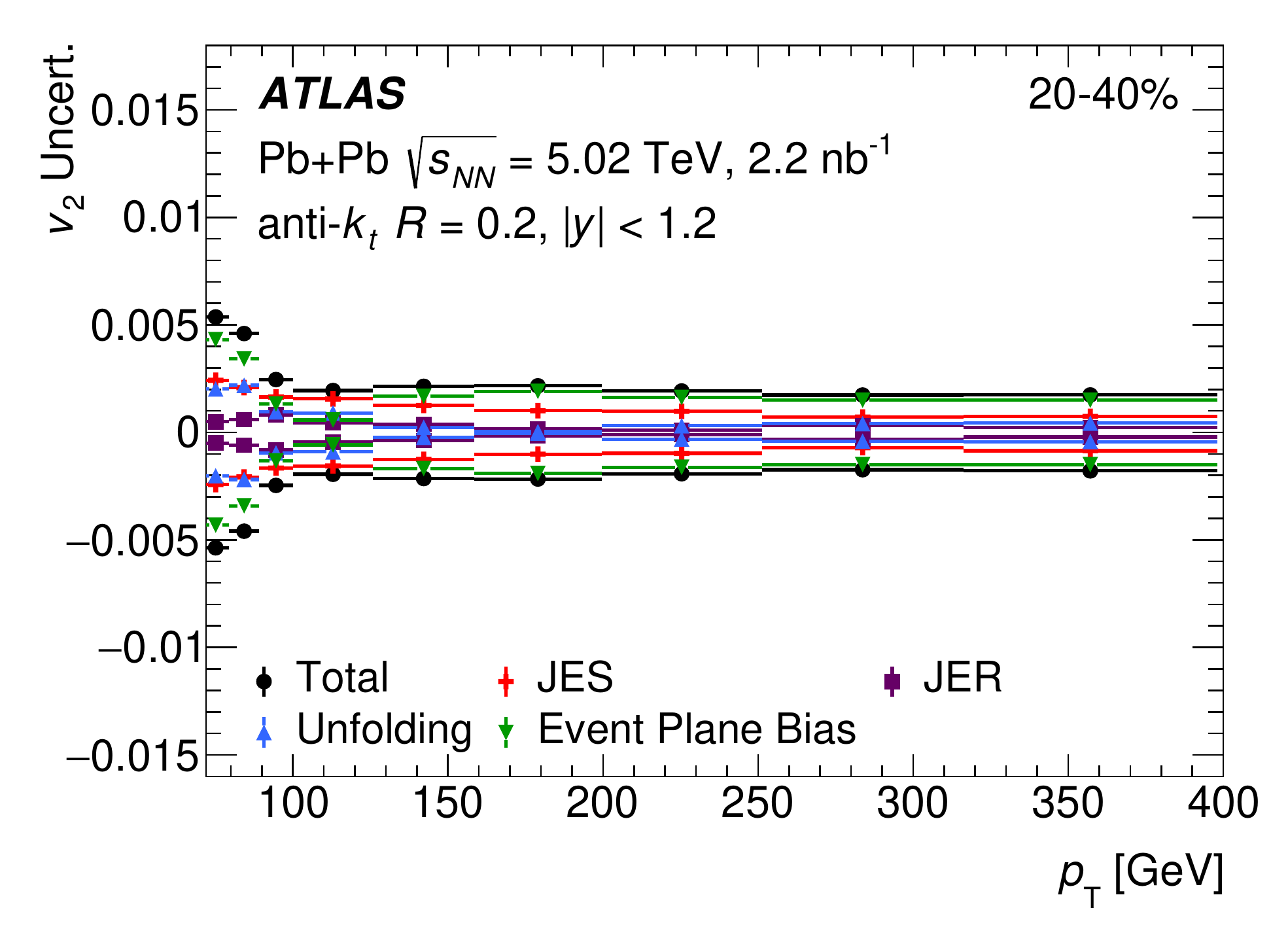}
}
\subfloat{
\centering
\includegraphics[width=0.5\textwidth]{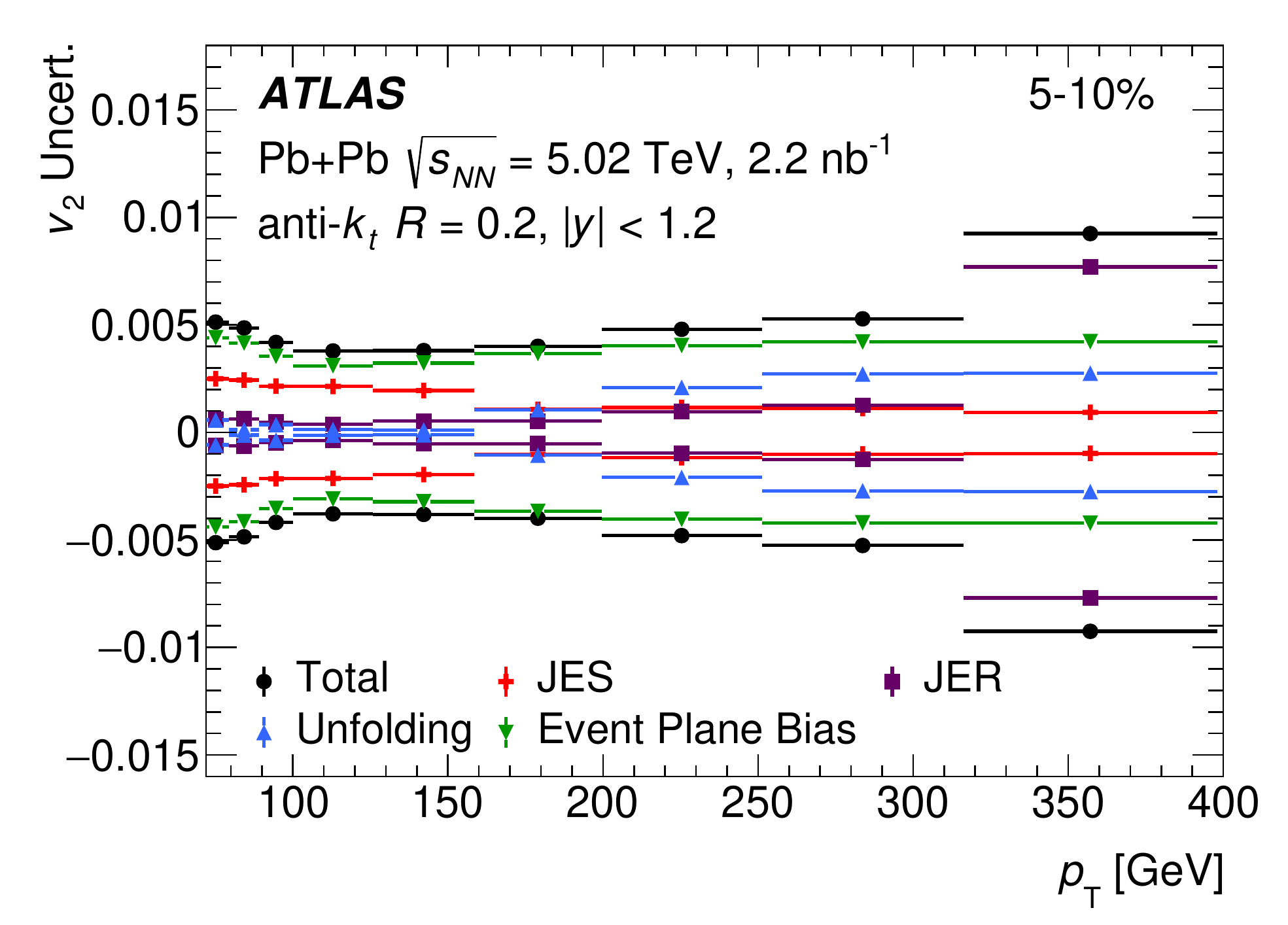}
}
\hfill
\subfloat{
\centering
\includegraphics[width=0.5\textwidth]{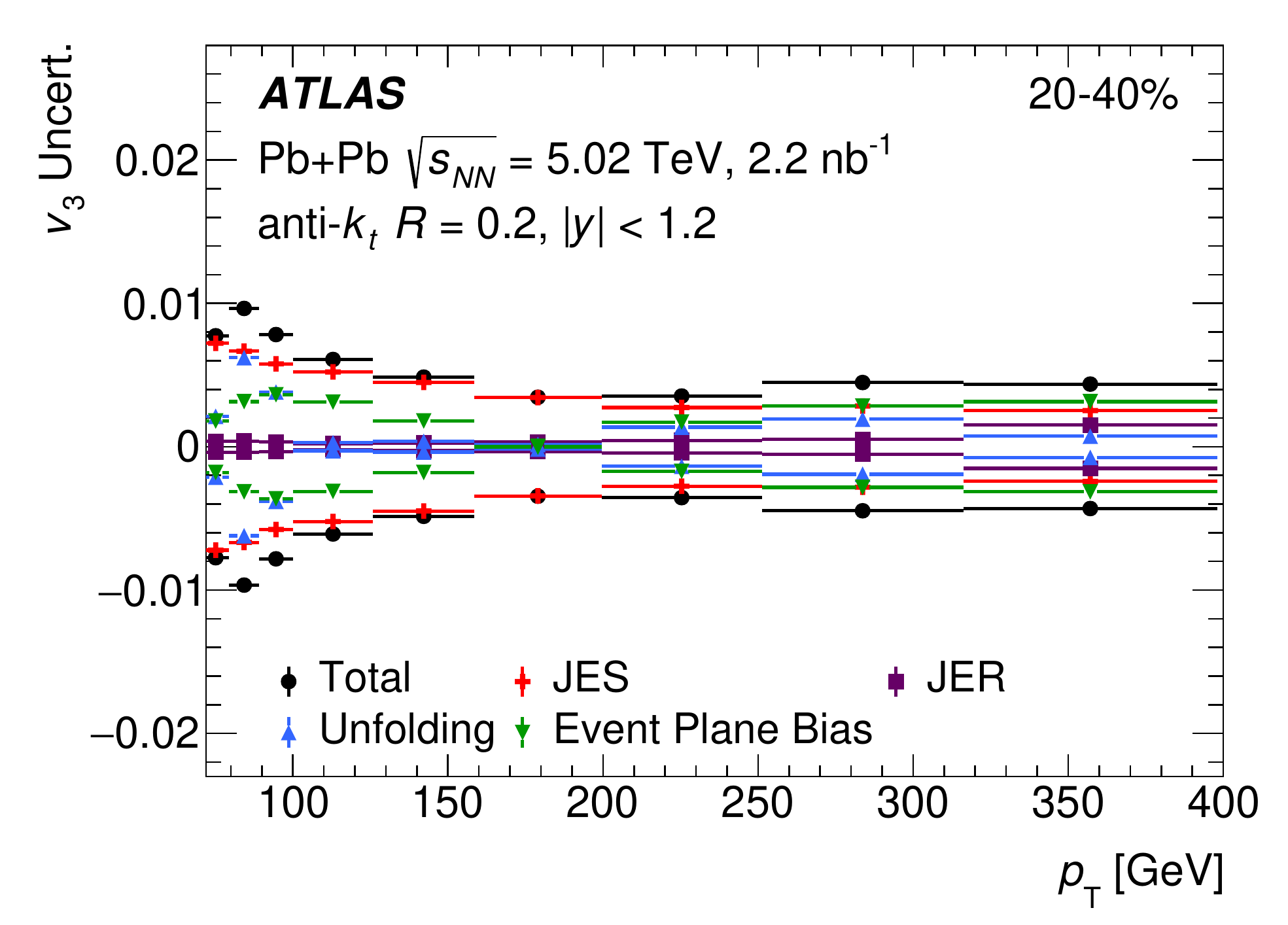}
}
\subfloat{
\centering
\includegraphics[width=0.5\textwidth]{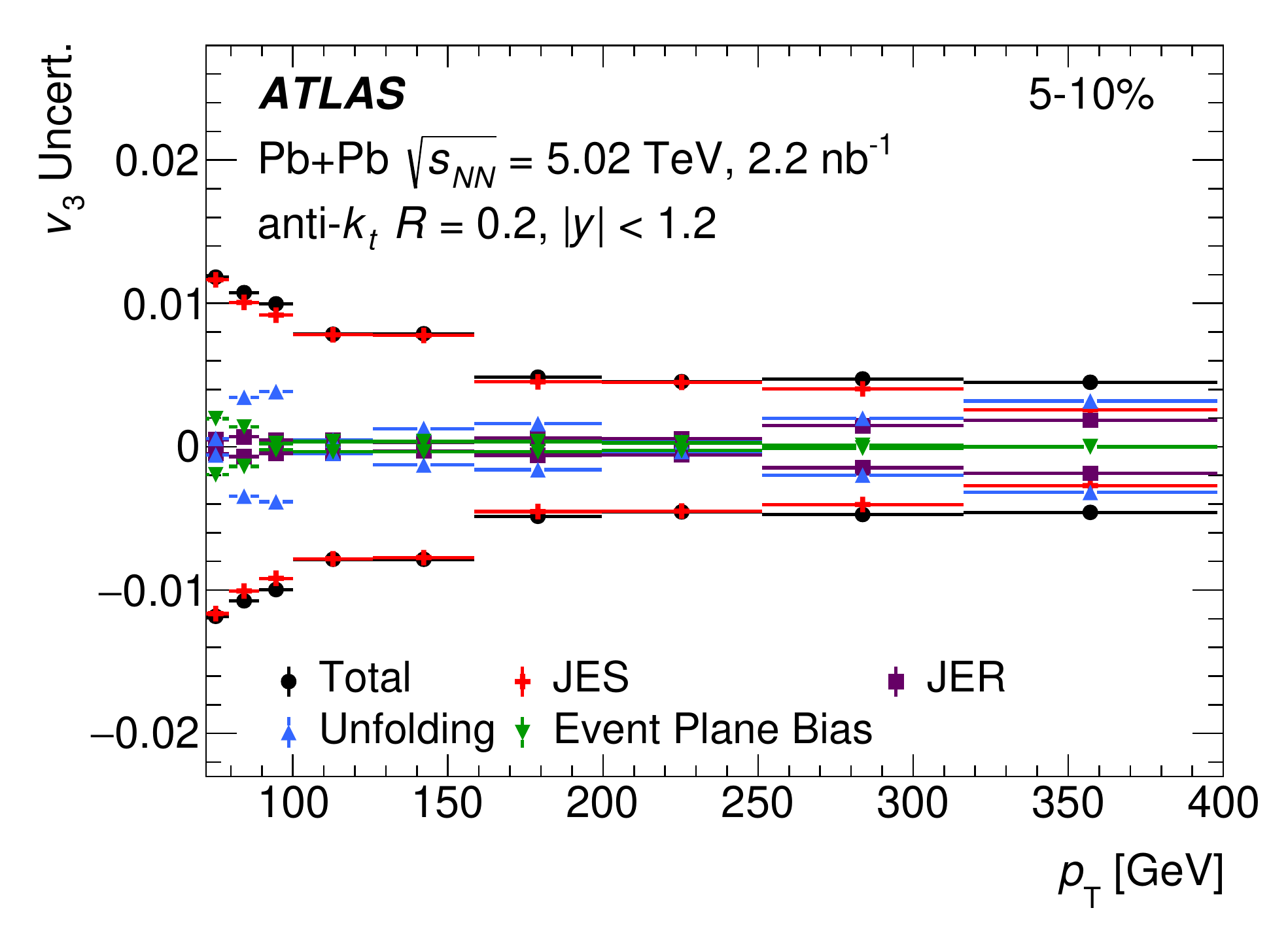}
}
\hfill
\subfloat{
\centering
\includegraphics[width=0.5\textwidth]{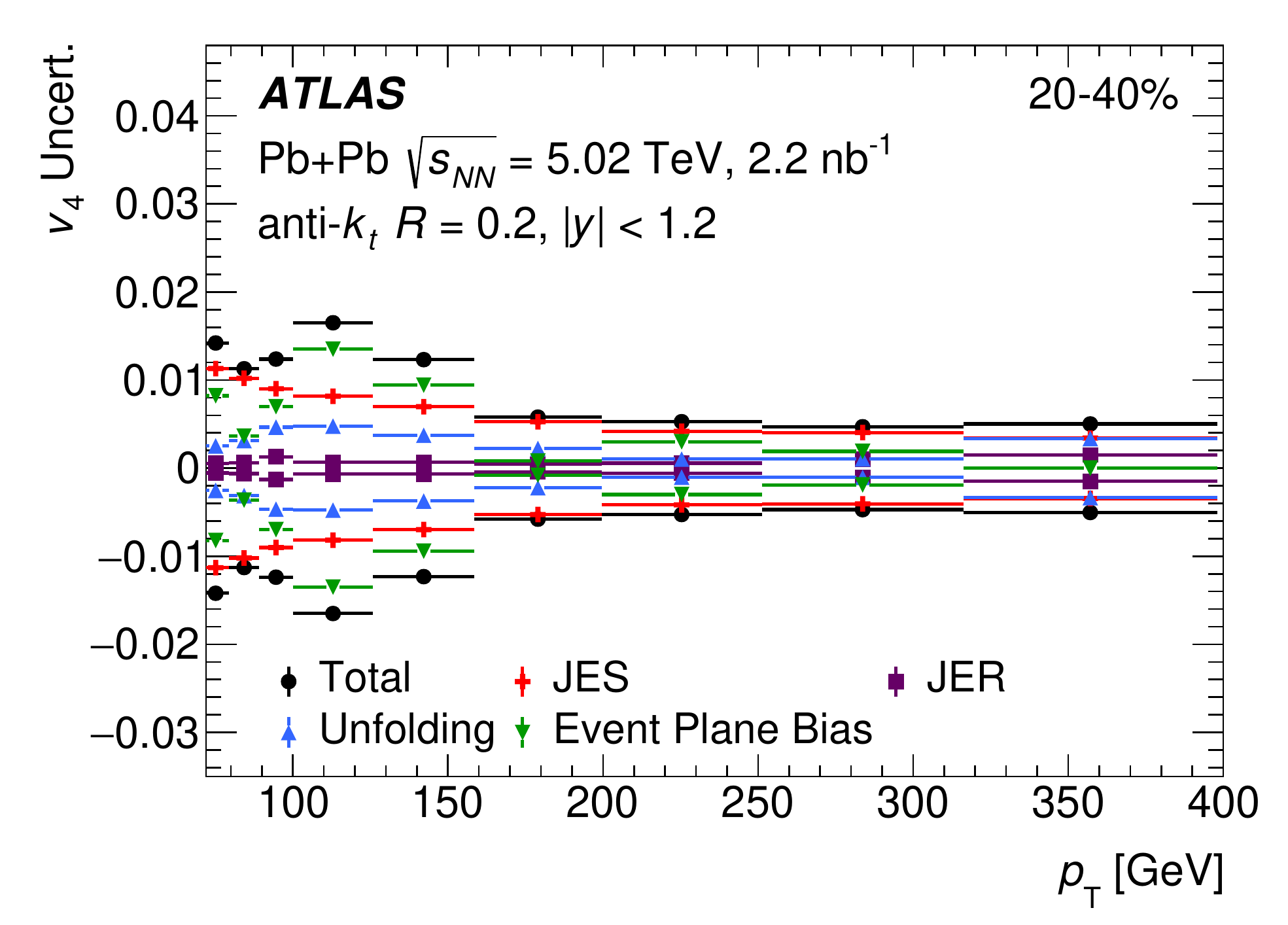}
}
\subfloat{
\centering
\includegraphics[width=0.5\textwidth]{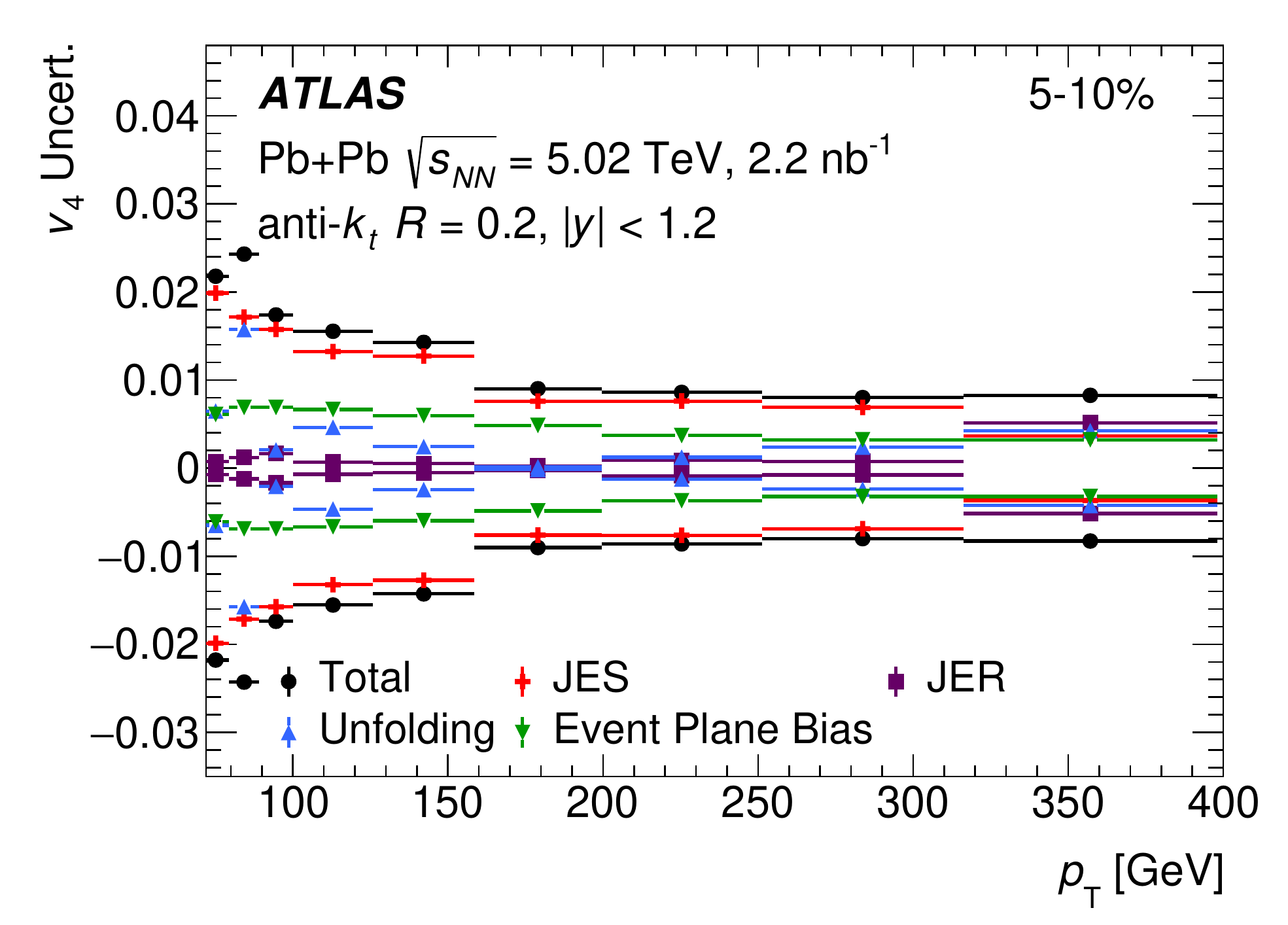}
}
\caption{The systematic uncertainties in \vtwo\ (top),
\vthree\ (middle), and \vfour\ (bottom) for 20--40\% (left)
and 5--10\% (right) centrality \pbpb\ collisions as a function of \pt. Each panel shows the total systematic uncertainty
as well as the size of the uncertainty from each of the sources, namely the JES, JER, unfolding, and event-plane bias.}
\label{fig:sysTotal}
\end{figure*}

The total systematic uncertainties of the \vn\ values and the contributions from each source are summarized in
Figure~\ref{fig:sysTotal}.
The largest uncertainty for \vtwo\ is the event-plane bias uncertainty, while for \vthree\ and \vfour\ the uncertainty in the \dphin\ dependence of the JES is largest.
The bin-to-bin variations in the unfolding uncertainty are largely statistical in nature.
 
\begin{figure*}[ht]
\centering
\subfloat{
\centering
\includegraphics[width=0.5\textwidth]{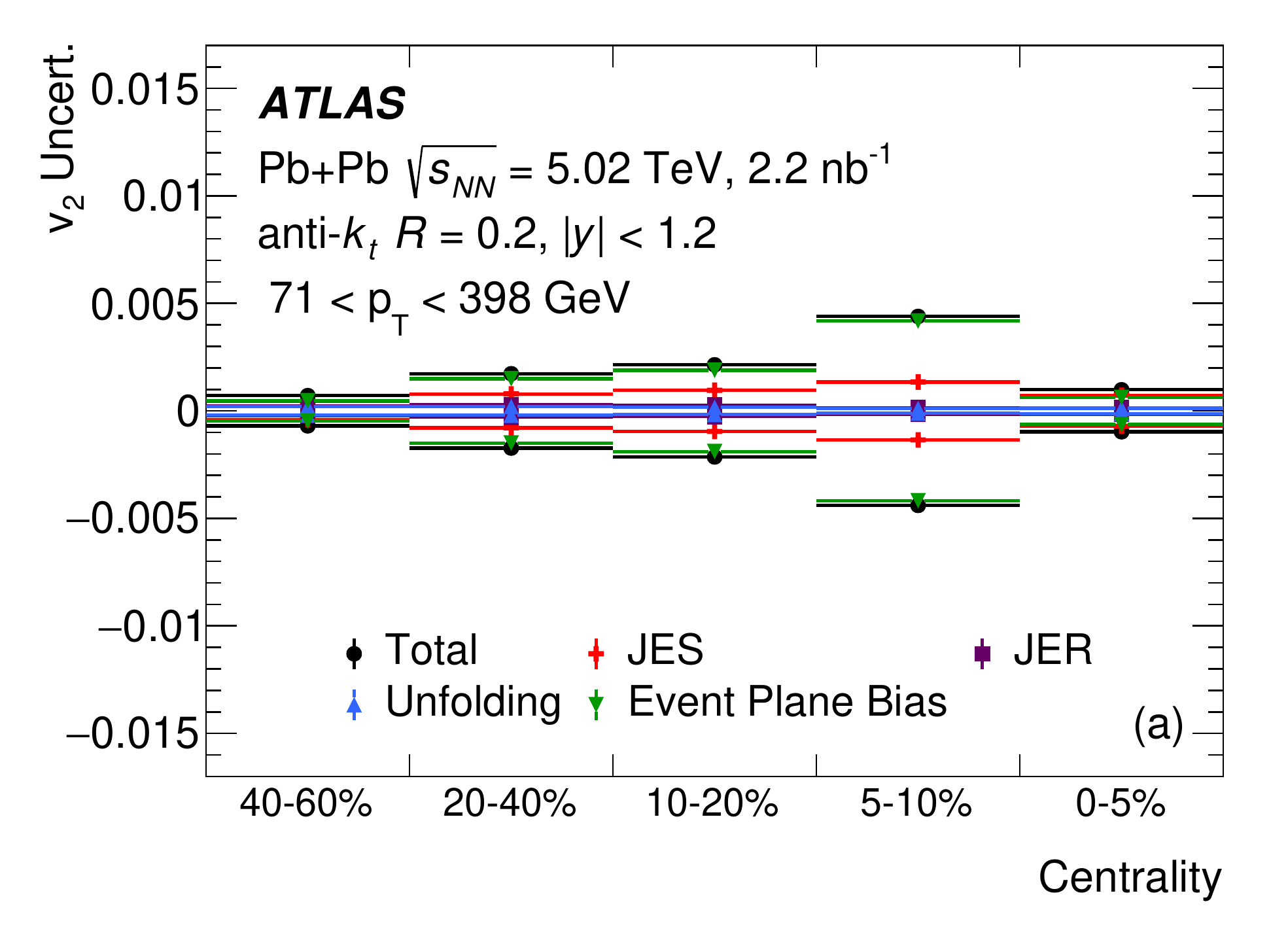}
}
\subfloat{
\centering
\includegraphics[width=0.5\textwidth]{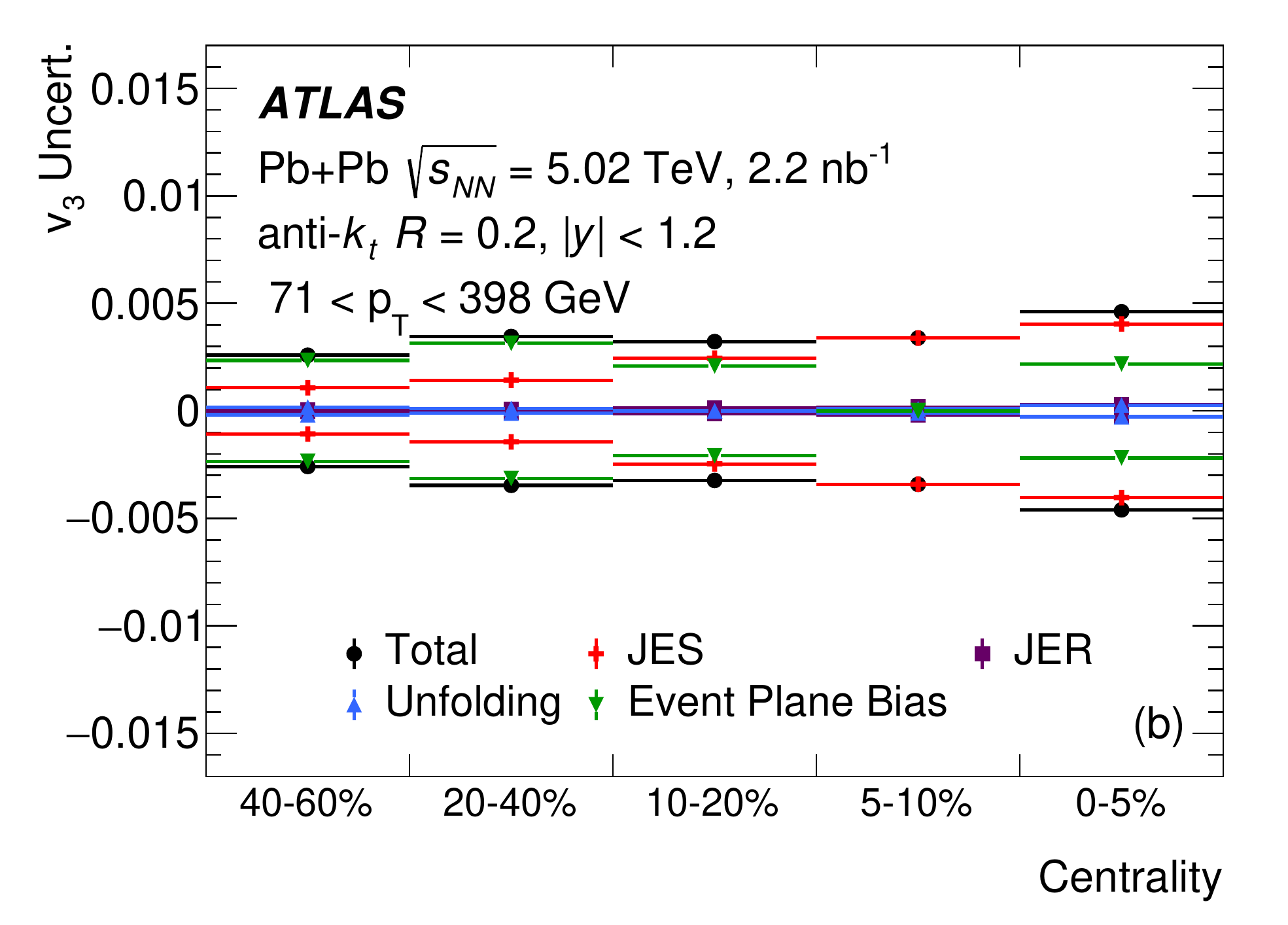}
}
\hfill
\subfloat{
\centering
\includegraphics[width=0.5\textwidth]{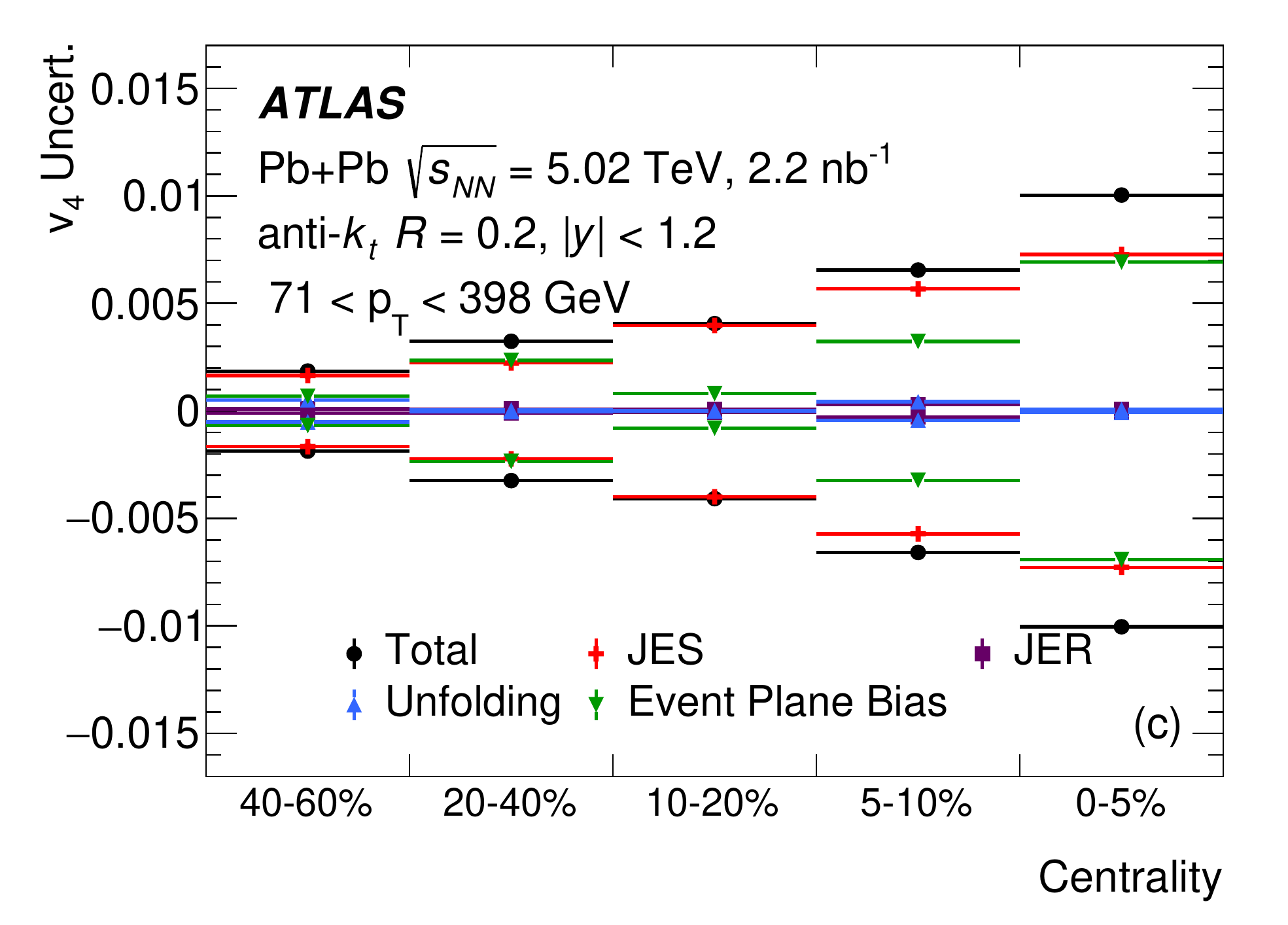}
}
\caption{The systematic uncertainties in (a) \vtwo,
(b) \vthree, and (c) \vfour\ for $\pt = \text{71--398}$~\GeV\ jets as a function of centrality.
Each panel shows the total systematic uncertainty as well as the size of the uncertainty from each of the sources, namely the JES, JER, unfolding, and event-plane bias.}
\label{fig:sysIncl}
\end{figure*}

Figure~\ref{fig:sysIncl} shows the systematic uncertainties of the \vn\ values measured in the inclusive \pt\ bin of 71--398~\GeV. The JES and JER uncertainties are smaller than those in
the \pt\ differential measurements as the variations largely move the jets within the inclusive \pt\ bin. Similarly, the unfolding uncertainty becomes smaller as the unfolding is
a smaller effect for the inclusive bin. The event-plane bias is the largest uncertainty in \vtwo, while the JES and event-plane bias are largest uncertainties in  \vthree\ and \vfour.
 
\FloatBarrier
 
\section{Results}
 
\begin{figure*}[ht]
\centering
\subfloat{
\centering
\includegraphics[width=0.5\textwidth]{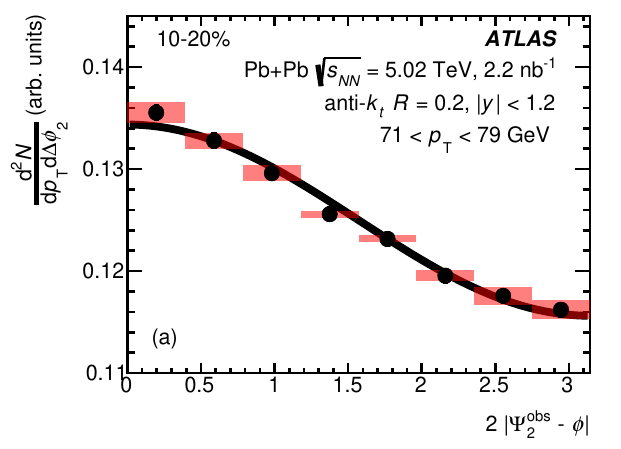}
}
\subfloat{
\centering
\includegraphics[width=0.5\textwidth]{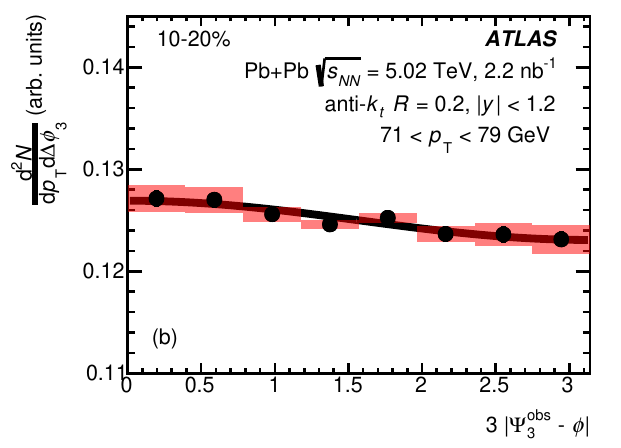}
}
\hfill
\subfloat{
\centering
\includegraphics[width=0.5\textwidth]{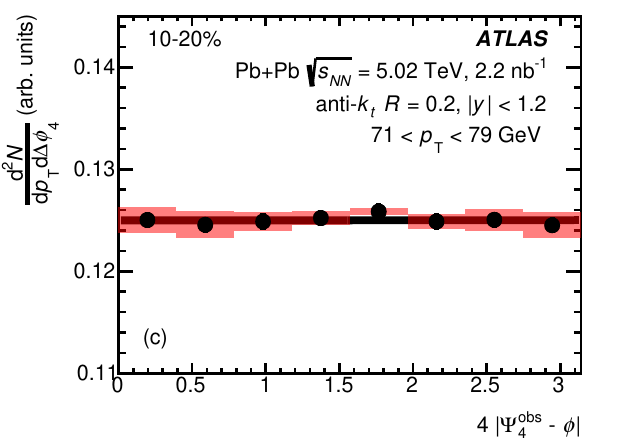}
}
\caption{Angular distribution of jets with respect to the
(a) $\Psi_2^\mathrm{obs}$, (b) $\Psi_3^\mathrm{obs}$, and (c) $\Psi_4^\mathrm{obs}$ planes, $n|\Psi_n^\mathrm{obs}-\phi|$, for jets with $71 < \pt < 79$~\GeV\ in the 10--20\% centrality bin.
The error bars show the statistical uncertainties, which are small compared to the size of the data points, and the boxes show the
systematic uncertainties. The black curve shows a fit of the data points to the function $A(1+2 v_{n}^\mathrm{obs} \cos (n\Delta \phi_n))$.}
\label{fig:dphi23}
\end{figure*}

Figure~\ref{fig:dphi23} shows an example of the angular distribution of jets with
respect to the \Psitwo, \Psithree, and \Psifour\ planes,
for jets with $71< \pt < 79$~\GeV\ in the 10--20\% centrality bin.  For both the \Psitwo\ and \Psithree\ dependence there are more jets
in-plane than out-of-plane, although for \Psithree\ the angular dependence is smaller. There is no significant dependence of the jet yield on the angle
with respect to \Psifour.
 
\begin{figure}[ht]
\centering
\includegraphics[width=0.8\textwidth]{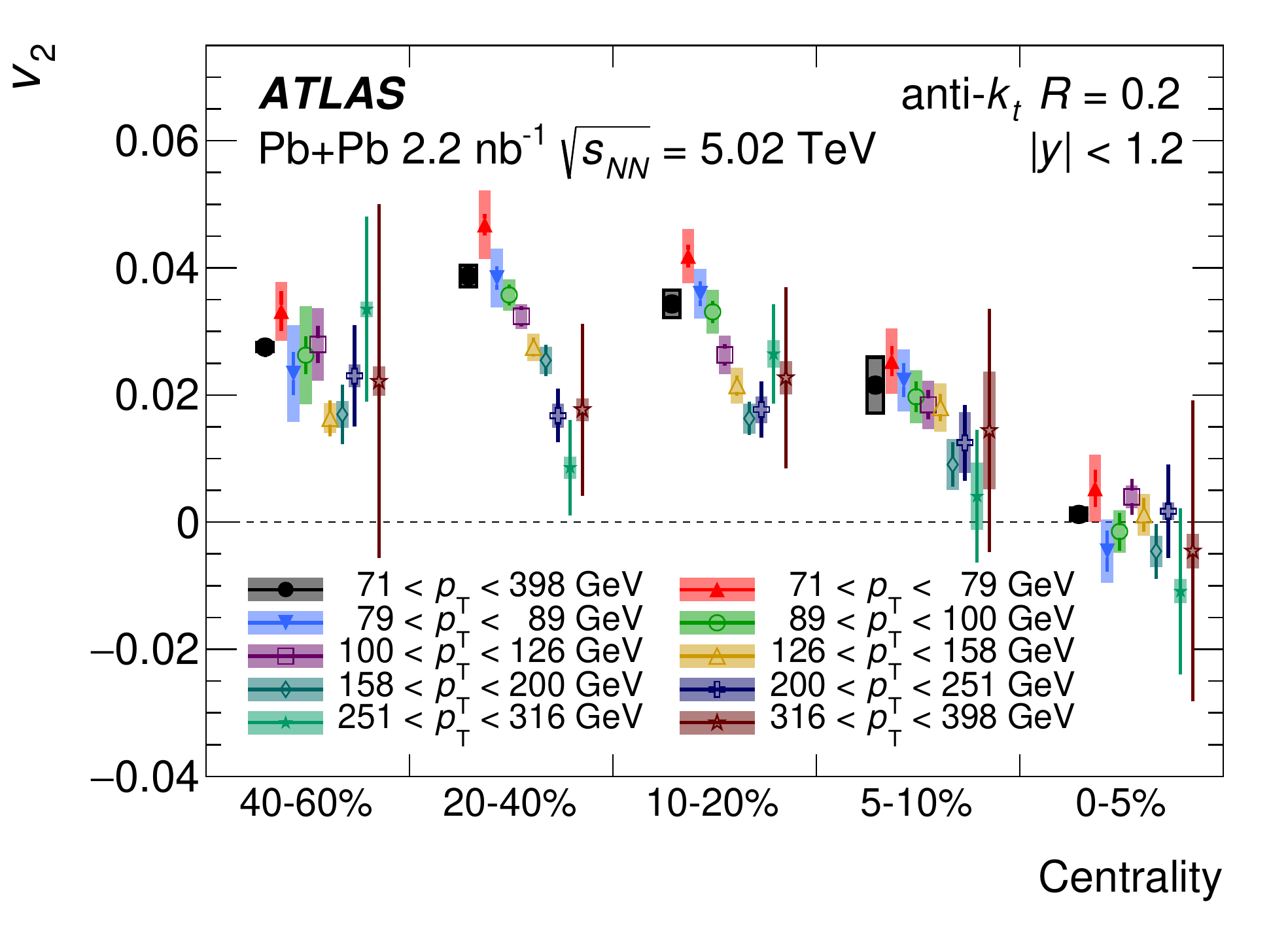}
\caption{The \vtwo\ values for
$R = 0.2$ jets as a function of centrality for jets in several \pt\ ranges, as indicated in the legend. The error bars represent the statistical error from the fits,
while the error boxes represent the systematic uncertainties.}
\label{fig:resultsv2}
\end{figure}

\begin{figure}[ht]
\centering
\includegraphics[width=0.8\textwidth]{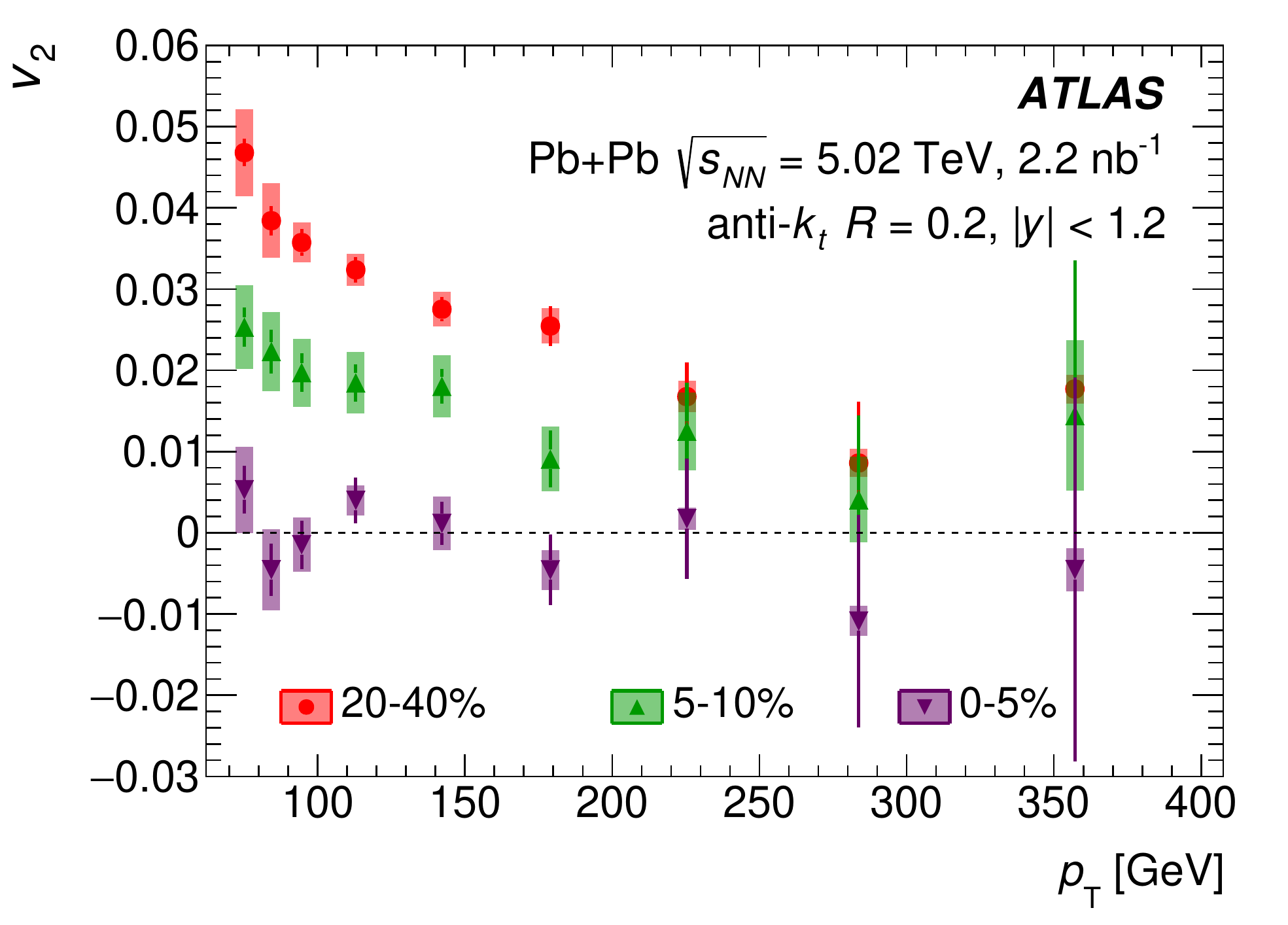}
\caption{The \vtwo\ values for
$R = 0.2$ jets as a function of \pt\ for 0--5\%, 5--10\%, and 20--40\% centrality collisions. The error bars represent the statistical error from the fits,
while the error boxes represent the systematic uncertainties.}
\label{fig:v2cent}
\end{figure}

The  $v_2$  values as a function of centrality for different \pt\ selections are shown in Figure~\ref{fig:resultsv2}.
The \vtwo\ values are consistent with zero in the most central collisions, and positive for all other centrality bins over the full \pt\ range.
For the  lower \pt\ ranges the \vtwo\ values are measured to be as large as 0.05 in mid-central collisions.
The \vtwo\ shows a decreasing trend with \pt\ in mid-central collisions, with a \vtwo\ of approximately 0.01--0.02 for jets with $\pt = \text{200--251}$~\GeV.
The value of \vtwo\ decreases for jets which have been shown in previous measurements to be less modified by the QGP, namely jets in peripheral collisions and high-\pt\ jets.
Figure~\ref{fig:v2cent} shows the \vtwo\ values for 0--5\%, 5--10\%, and 20--40\% centrality collisions as a function of jet \pt. The value of \vtwo\ decreases from
the more peripheral 20--40\% collisions to the
more central collisions, where the path-length difference between in-plane and out-of-plane is the smallest.
The dependence of the \vtwo\ on \pt\ in 5--10\% and 20--40\% collisions shows qualitatively similar behavior.
 
\begin{figure}[ht]
\centering
\includegraphics[width=0.8\textwidth]{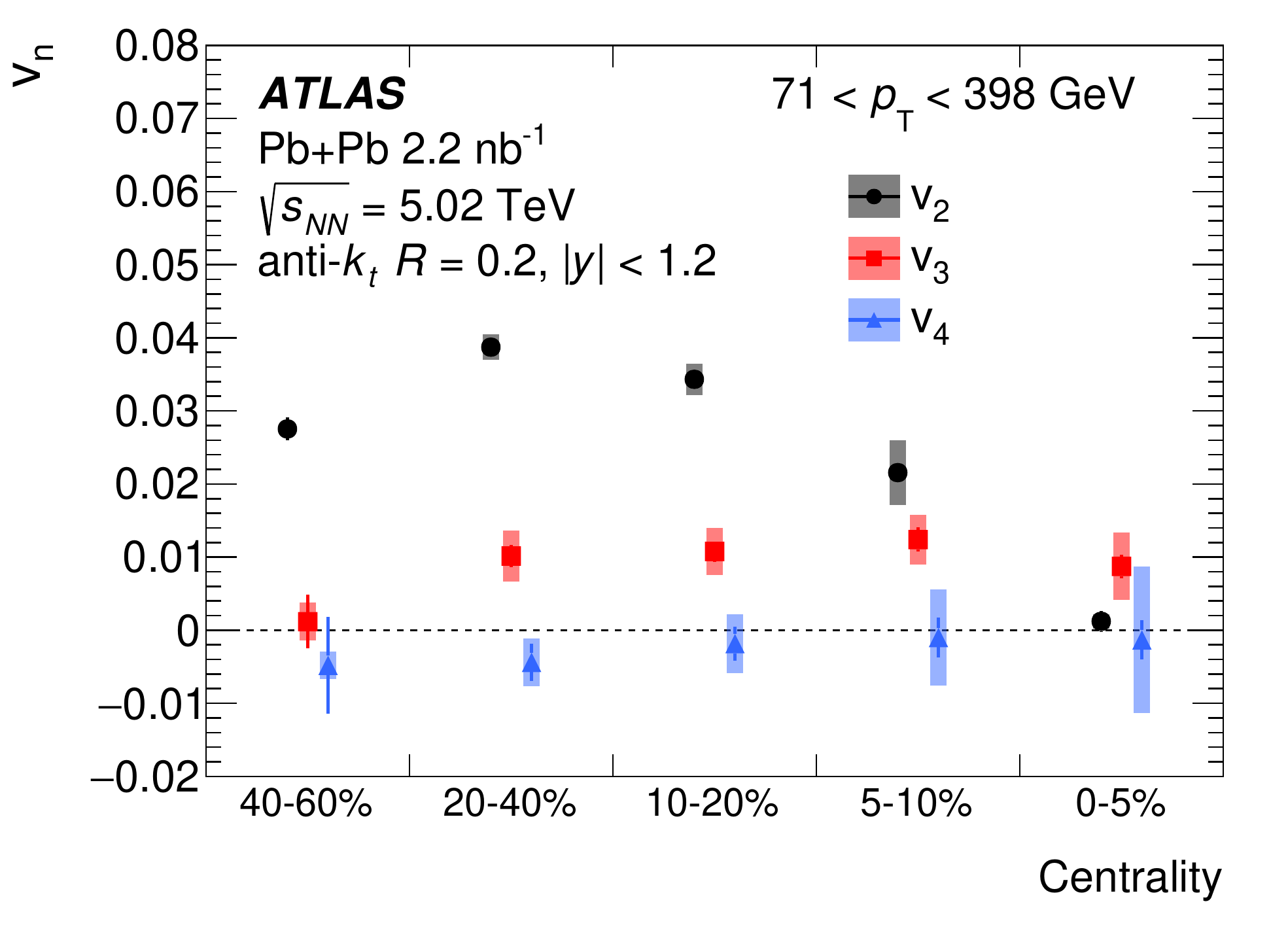}
\caption{The \vtwo, \vthree, and \vfour\ as a function of centrality for jets with $\pt = \text{71--398}$~\GeV. The error bars represent the statistical error from the fits,
while the error boxes represent the systematic uncertainties.}
\label{fig:ptinclusive}
\end{figure}

The centrality dependence for the \vtwo, \vthree, and \vfour\ is shown in Figure~\ref{fig:ptinclusive} for the full \pt\ range of the measurement, 71--398~\GeV.
The \vtwo\ is nonzero for jets with $\pt < 251$~\GeV\ in all but the most central collisions. The \vthree\ is positive and on the order of 0.01 for central and mid-central collisions, and consistent
with zero in the most peripheral collisions. The difference of the \vthree\ from 0 is 2.7$\sigma$ for 20--40\%, 3.1$\sigma$ for 10--20\%, 3.3$\sigma$ for 5--10\%, and
1.8$\sigma$ for 0--5\% collisions, where $\sigma$ is the quadrature sum of the statistical and systematic uncertainties.
The value of \vfour\ is compatible with zero. The measurements of \vthree\ and \vfour\ set a limit on the possible impact of
initial-state fluctuations on parton energy loss.

\FloatBarrier
 
\begin{figure*}[ht]
\centering
\subfloat{
\centering
\includegraphics[width=0.5\textwidth]{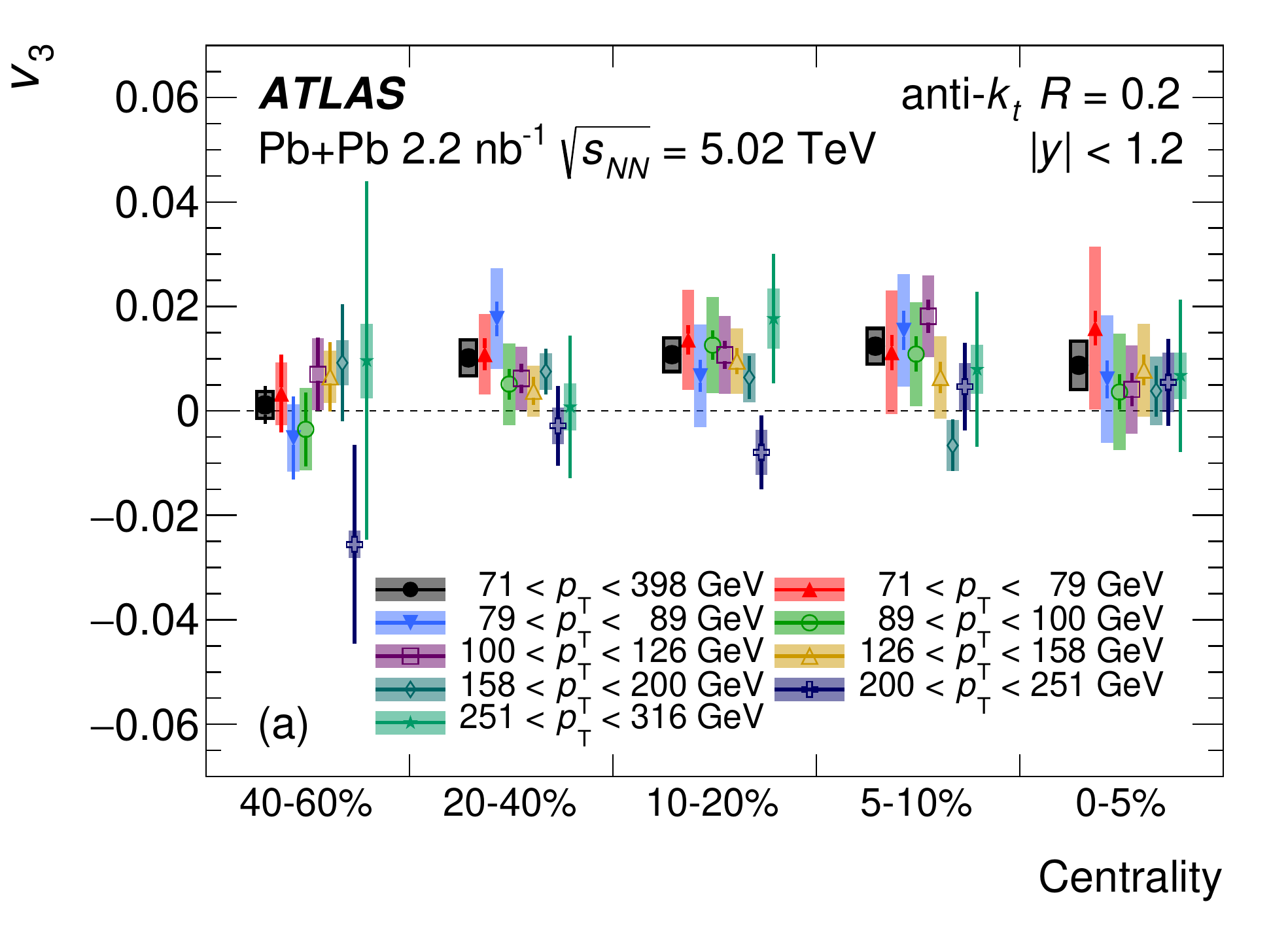}
}
\subfloat{
\centering
\includegraphics[width=0.5\textwidth]{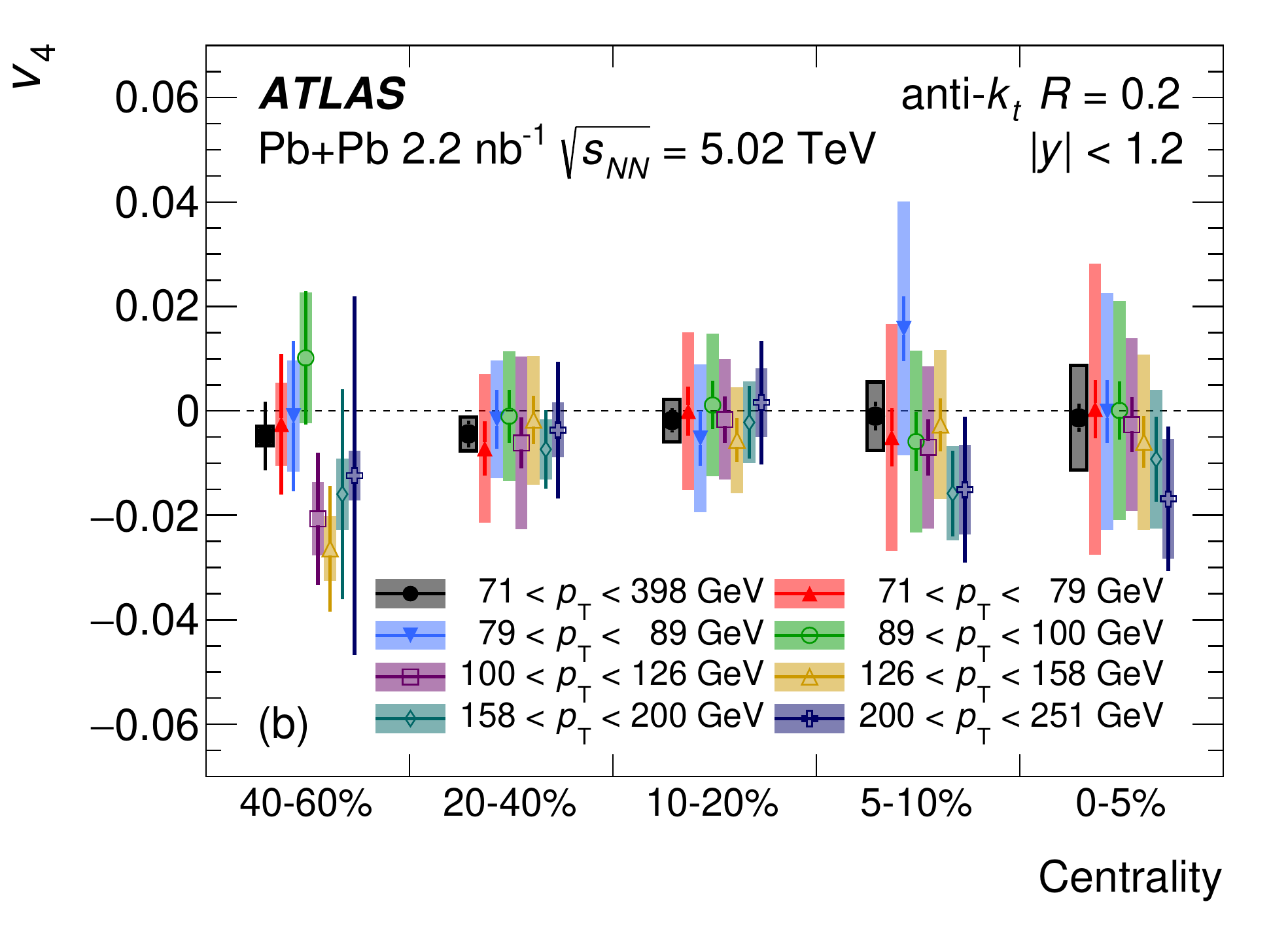}
}
\caption{The (a) $v_3$ and (b) $v_4$
of $R = 0.2$ jets as a function of centrality for jets in several \pt\ ranges, as indicated by the legend.
An inclusive bin of $\pt = \text{71--398}$~\GeV\ is also shown.
The error bars represent the statistical error from the fits,
while the error boxes represent the systematic uncertainties.}
\label{fig:resultsv34}
\end{figure*}

The centrality dependence of the measured \vthree\ and \vfour\ values for several \pt\ ranges are shown in Figure~\ref{fig:resultsv34}.
The \vthree\ shows no significant \pt\ or centrality dependence, with larger statistical and systematic uncertainties than in the measurement in the inclusive \pt\ bin.
The \vfour\ measurement is consistent with zero as a function of both \pt\ and centrality, with larger statistical uncertainties due to the poorer event-plane resolution
than that for the second- and third-order event-plane angles.
 
\FloatBarrier
 
\begin{figure*}[ht]
\centering
\subfloat{
\centering
\includegraphics[width=0.5\textwidth]{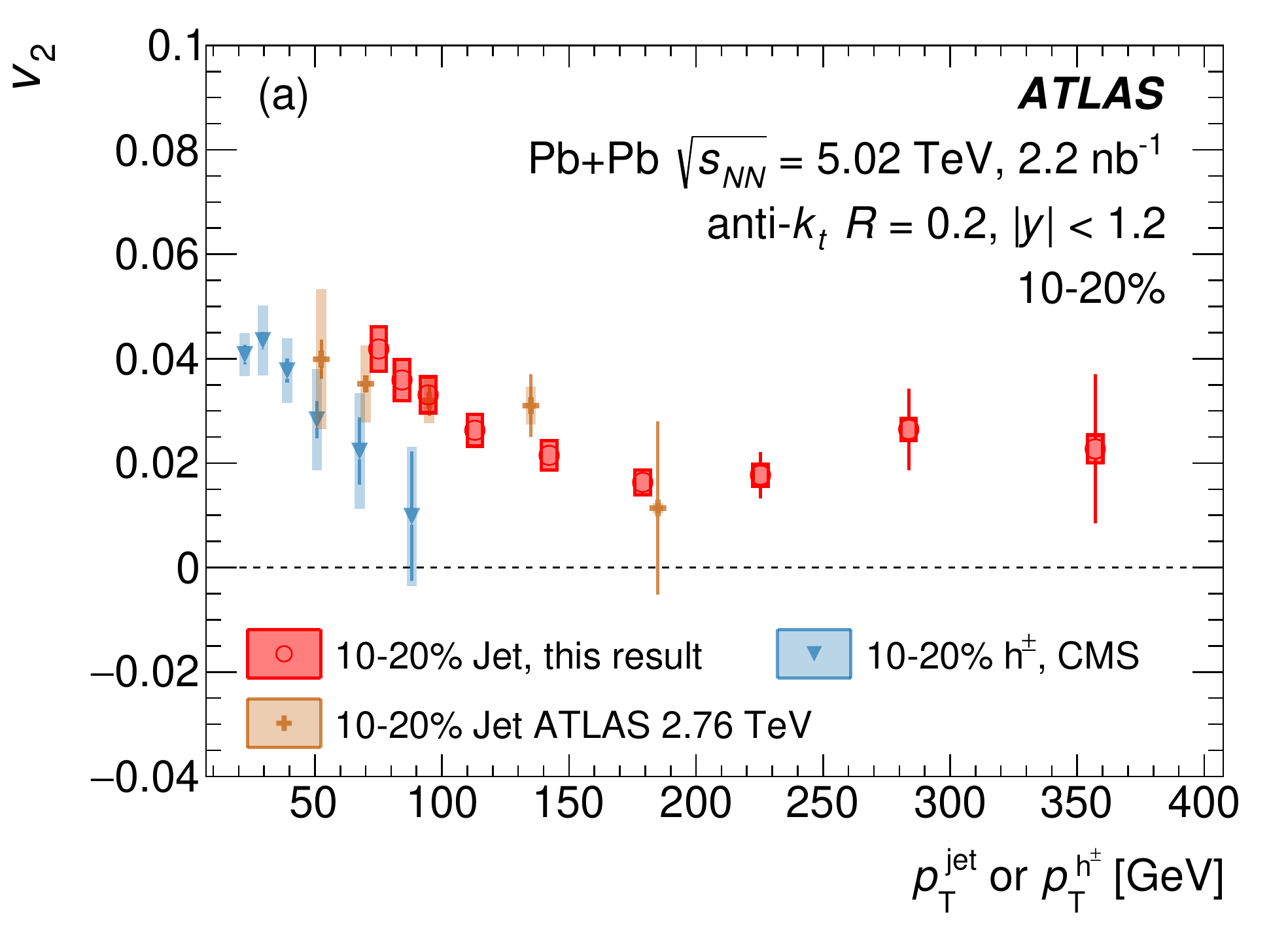}
}
\subfloat{
\centering
\includegraphics[width=0.5\textwidth]{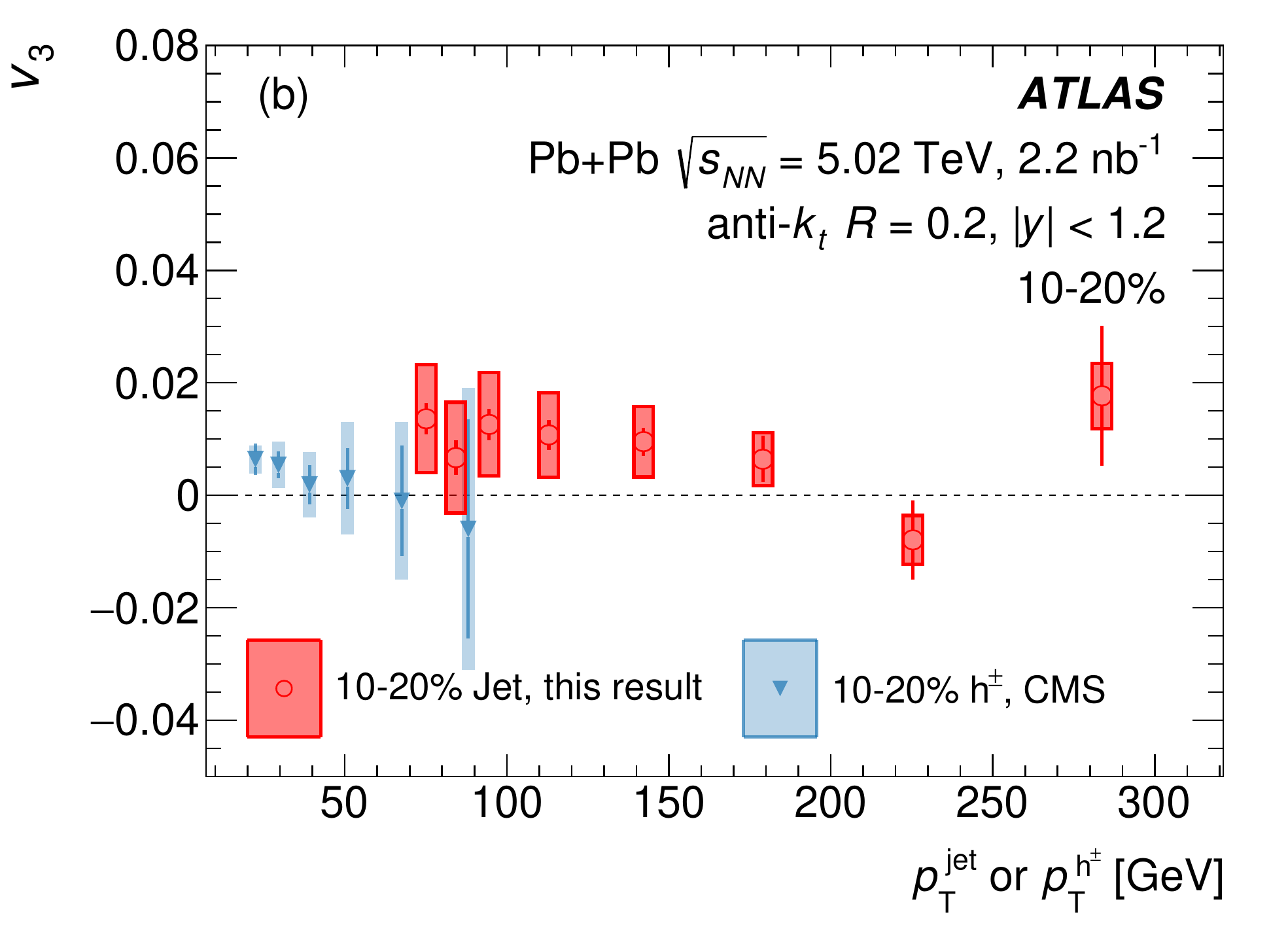}
}
\caption{The (a) \vtwo\ and (b) \vthree\ as a function of \pt\ for jets in 10--20\% centrality collisions in this measurement (red circles) compared with the \vtwo\ of jets in $\sqrtsnn = 2.76$~\TeV\
\PbPb\ collisions from Ref.~\cite{Aad:2013sla} (brown crosses) and the \vtwo\ and \vthree\ of charged particles in $\sqrtsnn = 5.02$~\TeV\ \PbPb\ collisions from
Ref.~\cite{Sirunyan:2017pan} (blue triangles).}
\label{fig:CMSCompCent2}
\end{figure*}

\begin{figure*}[ht]
\centering
\subfloat{
\centering
\includegraphics[width=0.5\textwidth]{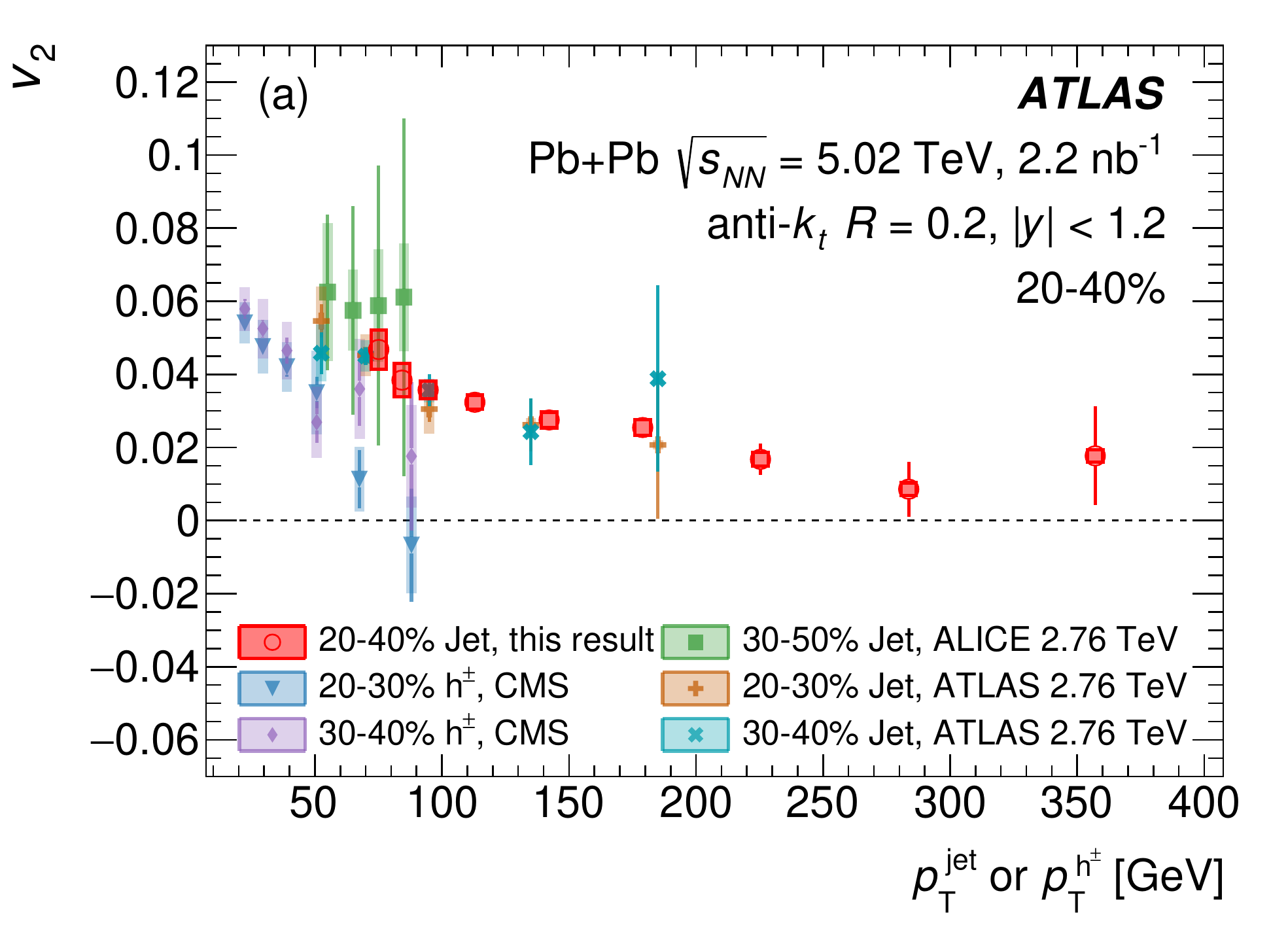}
}
\subfloat{
\centering
\includegraphics[width=0.5\textwidth]{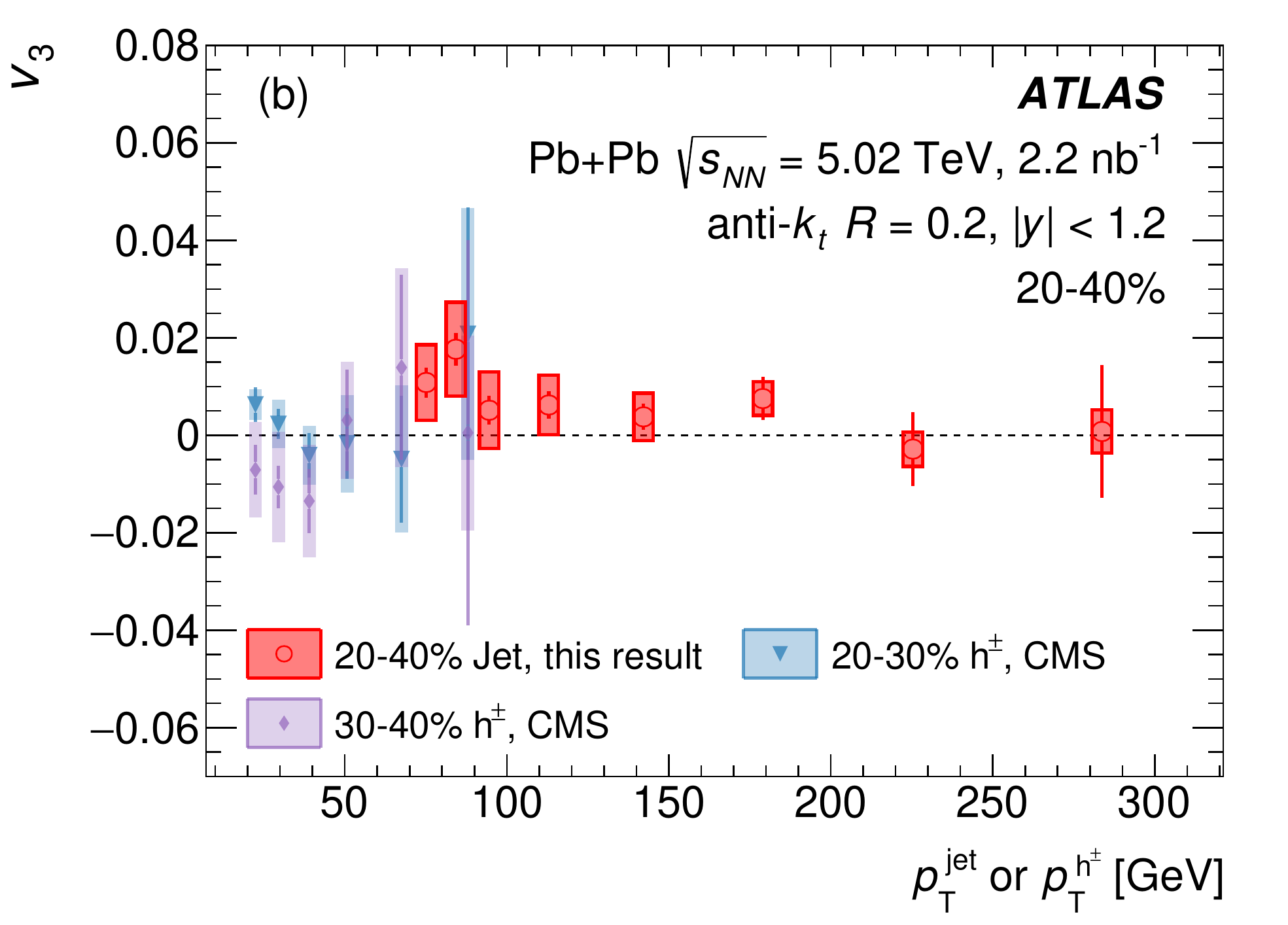}
}
\caption{The (a) \vtwo\ and (b) \vthree\ as a function of \pt\ for jets in 20--40\% centrality collisions in this measurement (red circles)
compared with the \vtwo\ of jets in $\sqrtsnn = 2.76$~\TeV\
\PbPb\ collisions from Ref.~\cite{Aad:2013sla} for 20--30\% centrality collisions (brown crosses) and 30--40\% centrality collisions (cyan X markers)
and from Ref.~\cite{Adam_2016} for 30--50\% centrality collisions (green squares)
and the \vtwo\ and \vthree\ of charged particles in $\sqrtsnn = 5.02$~\TeV\ \PbPb\ collisions from
Ref.~\cite{Sirunyan:2017pan} for 20--30\% centrality collisions (blue triangles) and 30--40\% centrality collisions (purple diamonds).}
\label{fig:CMSCompCent1}
\end{figure*}

Figures~\ref{fig:CMSCompCent2} and \ref{fig:CMSCompCent1} compare the results of this  measurement with the jet \vtwo\ measurements at $\sqrtsnn = 2.76$~\TeV\
for fully reconstructed jets from Ref.~\cite{Aad:2013sla} and charged-particle jets from Ref.~\cite{Adam_2016}
for the 10--20\% and 20--40\% centrality bins. The measurement shows good agreement with the previous results, with no significant evidence of
a dependence of the \vn\ values on the collision
energy. The \vtwo\ and \vthree\ of charged particles from Ref.~\cite{Sirunyan:2017pan} is also shown.
The results show a qualitatively similar \pt\ dependence, with the charged-particle \vn\ distribution shifted to lower \pt. This is consistent
with the expectation that high-\pt\ charged particles are likely produced from jets at a higher \pt.
This result improves on both the \pt\ reach and precision of previous measurements of high-\pt\ jet \vn.
 
\FloatBarrier
 
\begin{figure*}[ht]
\centering
\subfloat{
\centering
\includegraphics[width=0.5\textwidth]{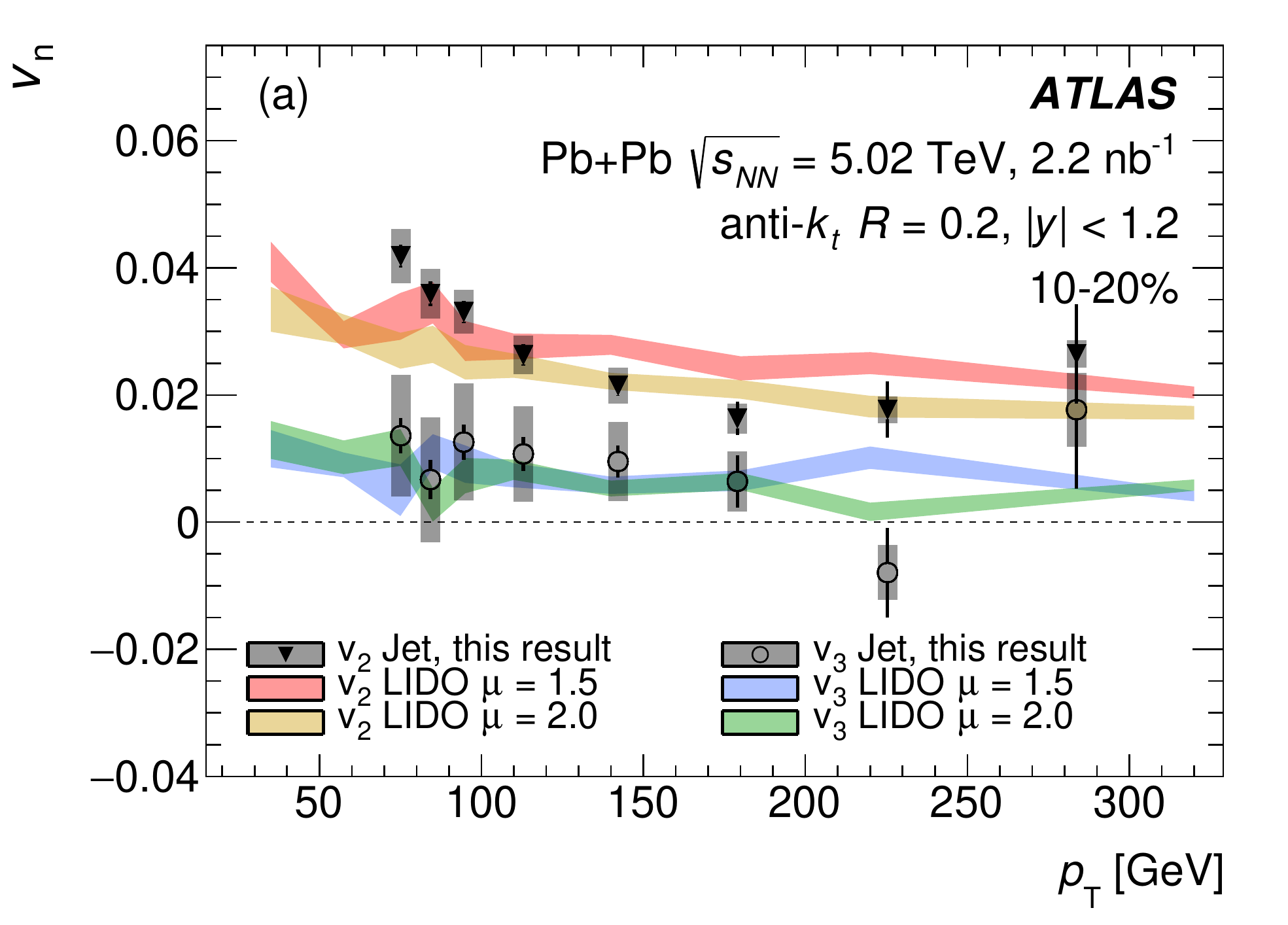}
}
\subfloat{
\centering
\includegraphics[width=0.5\textwidth]{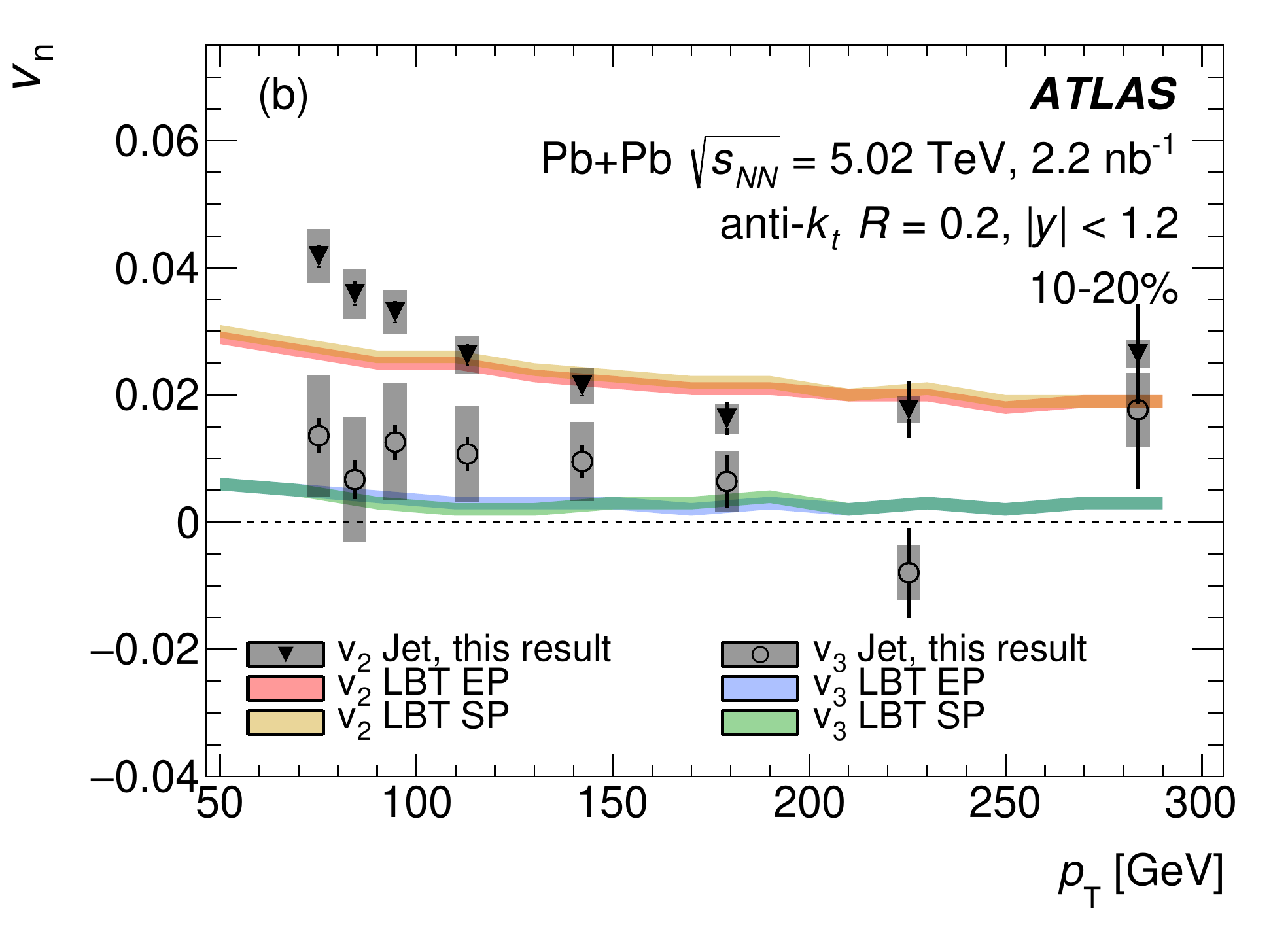}
}
\caption{\label{fig:theory} The \vtwo\ and \vthree\ for jets in 10--20\% centrality collisions compared with theoretical calculations using the (a) LIDO~\cite{ke2020qgp}  and (b) LBT~\cite{PhysRevC.91.054908,PhysRevC.99.054911,He:2018gxh} models. The LIDO calculation is shown for two values of the jet--medium coupling cutoff parameter, $\mu$. The LBT calculation is shown using the
event-plane method (EP) and the scalar-product method (SP).}
\end{figure*}

Figure~\ref{fig:theory} shows the \vtwo\ and \vthree\ as a function of \pt\ compared with theoretical calculations: LIDO from Ref.~\cite{ke2020qgp}
and the Linear Boltzmann Transport (LBT) model from Refs.~\cite{PhysRevC.91.054908,PhysRevC.99.054911,He:2018gxh}.
LIDO is a transport model including both elastic jet--medium collisions and medium-induced radiative processes, as well as a simple model for the response of the medium.
The \vn\ is computed with an event-by-event model of the QGP medium~\cite{bernhard19}.
The calculations are performed using two values for the jet--medium coupling cutoff parameter: $\mu = 1.5$ and $\mu = 2.0$. This parameter is related to the
strength of the coupling between the jet and the medium, where a smaller $\mu$ value corresponds to a larger coupling~\cite{ke2020qgp}.
The choice of $\mu$ values is motivated by comparisons with measurements of jet quenching.
The LIDO model describes the \vtwo\ and \vthree\ well,
with the data favoring the $\mu = 2.0$ calculation at higher \pt\ and the $\mu = 1.5$ calculation at lower \pT.
The LBT model simulates the propagation of jet shower and thermal recoil partons in the same framework and includes
the effect of the jet-induced medium particles in the reconstruction of the final jets.
The model uses event-by-event hydrodynamics as described in Ref.~\cite{PhysRevC.97.064918}.
The calculation is shown using the event-plane method (EP), which does not include soft hadron fluctuations, and
the scalar-product method (SP), which does include soft hadron fluctuations~\cite{PhysRevC.66.034904}.
The LBT model agrees with the size of the \vthree\ within the uncertainties of this measurement,
but does not describe the variation of \vtwo\ as a function of \pT.

\FloatBarrier
 
\begin{figure*}[ht]
\centering
\subfloat{
\centering
\includegraphics[width=0.5\textwidth]{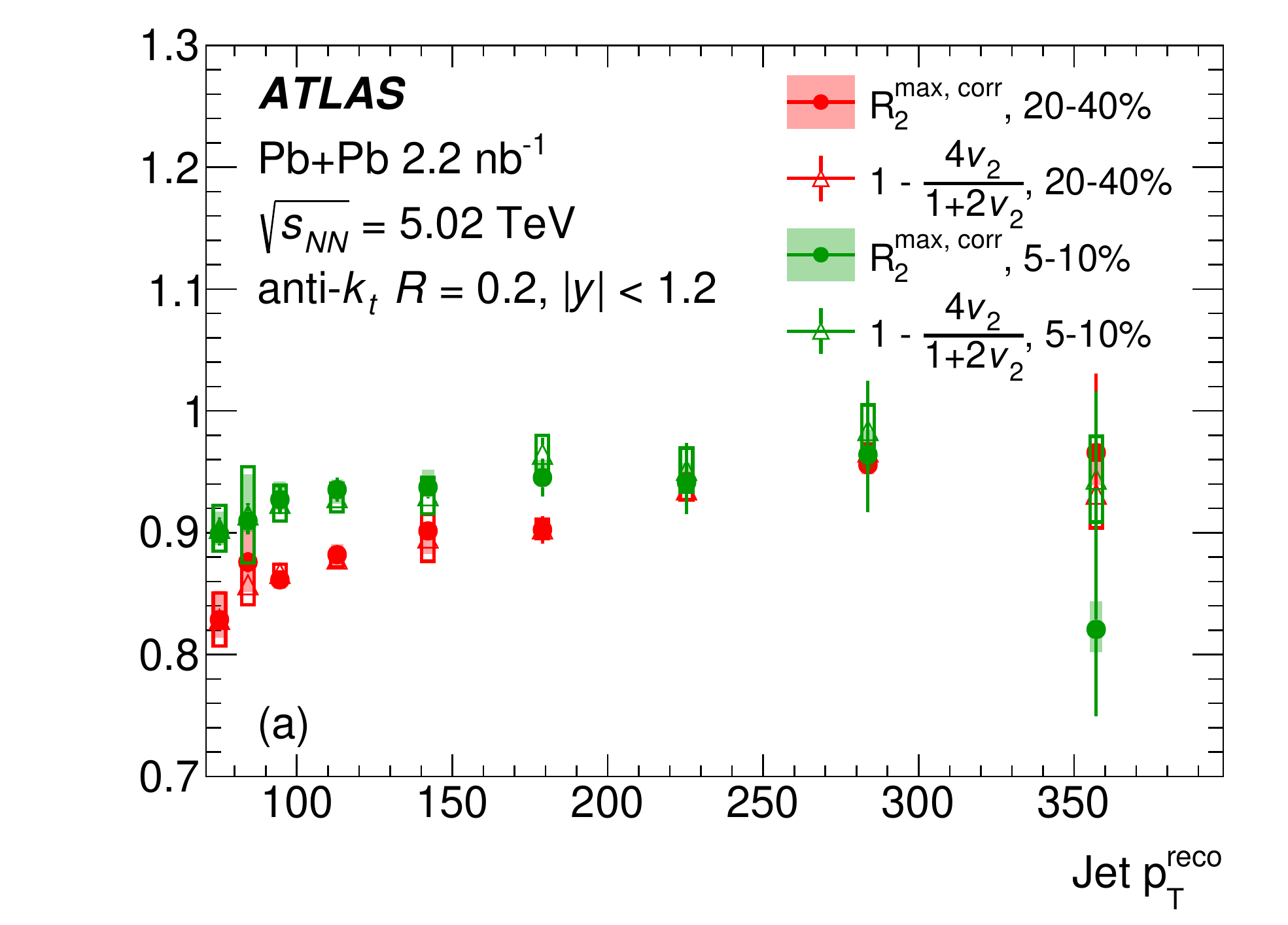}
}
\subfloat{
\centering
\includegraphics[width=0.5\textwidth]{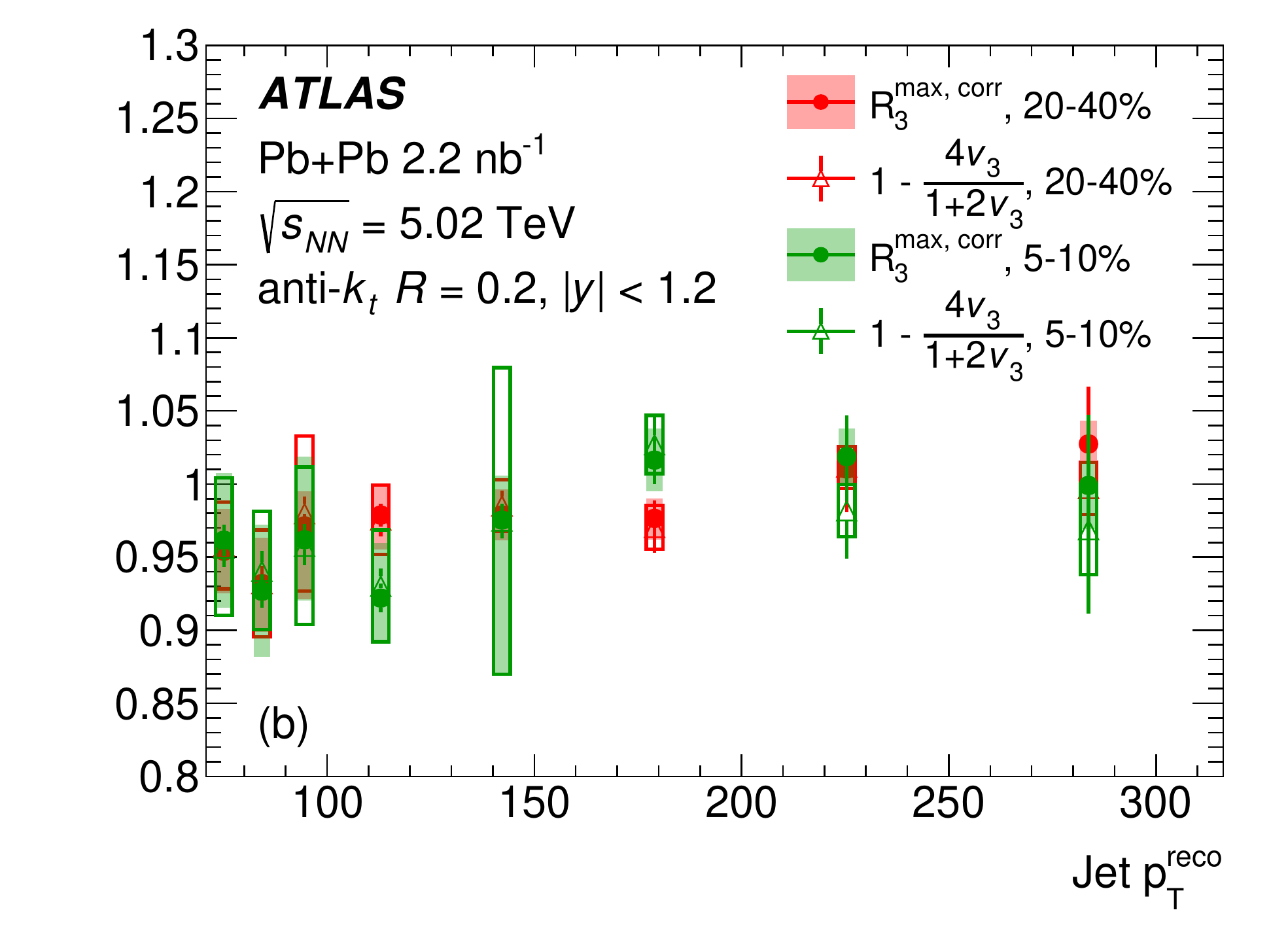}
}
\caption{\label{fig:ratio} \Rnmax\ for (a) $n=2$ and (b) $n=3$ as a function of \pT\ (filled circles).
Also shown is  $1 - 4\vn/(1+2\vn)$ (open circles).  The values are shown for 5--10\% (green) and 20--40\%
(red) collisions.}
 
\end{figure*}

It is interesting to compare the actual jet yields in-plane versus out-of-plane to study the
angular distribution of jets without imposing the $\cos(n \Delta \phi_n )$ shape modulation on the data.
The ratio of the jet yields in the most in-plane bin, $n \Delta \phi_n < \pi / 8$, to the most
out-of-plane bin, $n \Delta \phi_n > 7 \pi / 8$, is constructed:
\begin{equation}\nonumber
R_{n}^{\mathrm{max}} \equiv {\frac{d^2 N}{d\pT d\Delta \phi_n}\bigg|_{n \Delta \phi_n > 7 \pi / 8}} \bigg/
{\frac{d^2 N}{d\pT d\Delta \phi_n}\bigg|_{n \Delta \phi_n < \pi / 8}}.
\end{equation}
These yields must be corrected for the finite event-plane resolution, which is done by
assuming that the variation in \dphin\ is dominated by the
$\cos(n \Delta \phi_n )$ modulation such that
\begin{equation}\nonumber
\frac{d^2N_{\mathrm{jet}}^{\mathrm{corr}}}{d\pT d\Delta \phi_n} = \frac{d^2N_{\mathrm{jet}}^{\mathrm{obs}}}{d\pT d\Delta \phi_n}
\left(\frac{1+2v_n\cos{n\Delta \phi_n}}{1+2v_{n}^{\mathrm{obs}}\cos{n\Delta \phi_n}}\right).
\end{equation}
The ratio is further corrected for the effects of the finite bin width by assuming a $\cos(n \Delta \phi_n )$ modulation within each bin, and
calculating the yields at $n \Delta \phi_n = 0$ and $n \Delta \phi_n = \pi$, and taking the ratio of these values.
A similar method was used in Ref.~\cite{Aad:2013sla}.
This ratio, for $n= 2$ and $3$, is shown in Figure~\ref{fig:ratio}.
A purely $\cos(n \Delta \phi_n )$ modulation would cause \Rnmax\ to be $1 - 4\vn/(1+2\vn)$ and these calculated values are
compared with the \Rnmax\ values.  No deviation from the $\cos(n \Delta \phi_n )$ modulation is observed.
 
\clearpage
\FloatBarrier

\section{Conclusion}
The azimuthal variation of jet quenching is measured in 2.2~nb$^{-1}$ of \pbpb\ collisions at 5.02~\TeV, using the event-plane method to
extract the \vn\ coefficients of jets. The data were collected with the ATLAS detector at the LHC.
The \vtwo\ is found to be consistent with zero in the most central collisions,
with values up to 0.05 in mid-central collisions, and decreasing with increasing \pt. A first measurement of \vthree\ and
\vfour\ of jets is presented. The value of \vthree\ is found to be significantly above zero in mid-central collisions, with a value
of approximately 0.01 for central and mid-central collisions.
The \pt-differential measurement of \vthree\ shows no significant \pt\ or centrality dependence, while
\vfour\ is everywhere consistent with zero. The measured values are consistent with previous measurements of jet and high-\pt\
hadron \vn, and improve on both the \pt\ reach and precision of these previous results.
The positive \vtwo\ values in all but the most azimuthally symmetric collisions show the relationship between collision geometry and parton energy loss.
Further, the measurements of \vthree\ and \vfour\ will help set limits on the impact of initial-state fluctuations on energy loss.
These measurements can be used to constrain models of the path-length dependence of jet quenching.
\clearpage

\section*{Acknowledgments}
 

We thank CERN for the very successful operation of the LHC, as well as the
support staff from our institutions without whom ATLAS could not be
operated efficiently.
 
We acknowledge the support of
ANPCyT, Argentina;
YerPhI, Armenia;
ARC, Australia;
BMWFW and FWF, Austria;
ANAS, Azerbaijan;
SSTC, Belarus;
CNPq and FAPESP, Brazil;
NSERC, NRC and CFI, Canada;
CERN;
ANID, Chile;
CAS, MOST and NSFC, China;
Minciencias, Colombia;
MEYS CR, Czech Republic;
DNRF and DNSRC, Denmark;
IN2P3-CNRS and CEA-DRF/IRFU, France;
SRNSFG, Georgia;
BMBF, HGF and MPG, Germany;
GSRI, Greece;
RGC and Hong Kong SAR, China;
ISF and Benoziyo Center, Israel;
INFN, Italy;
MEXT and JSPS, Japan;
CNRST, Morocco;
NWO, Netherlands;
RCN, Norway;
MEiN, Poland;
FCT, Portugal;
MNE/IFA, Romania;
JINR;
MES of Russia and NRC KI, Russian Federation;
MESTD, Serbia;
MSSR, Slovakia;
ARRS and MIZ\v{S}, Slovenia;
DSI/NRF, South Africa;
MICINN, Spain;
SRC and Wallenberg Foundation, Sweden;
SERI, SNSF and Cantons of Bern and Geneva, Switzerland;
MOST, Taiwan;
TAEK, Turkey;
STFC, United Kingdom;
DOE and NSF, United States of America.
In addition, individual groups and members have received support from
BCKDF, CANARIE, Compute Canada and CRC, Canada;
COST, ERC, ERDF, Horizon 2020 and Marie Sk{\l}odowska-Curie Actions, European Union;
Investissements d'Avenir Labex, Investissements d'Avenir Idex and ANR, France;
DFG and AvH Foundation, Germany;
Herakleitos, Thales and Aristeia programmes co-financed by EU-ESF and the Greek NSRF, Greece;
BSF-NSF and GIF, Israel;
Norwegian Financial Mechanism 2014-2021, Norway;
NCN and NAWA, Poland;
La Caixa Banking Foundation, CERCA Programme Generalitat de Catalunya and PROMETEO and GenT Programmes Generalitat Valenciana, Spain;
G\"{o}ran Gustafssons Stiftelse, Sweden;
The Royal Society and Leverhulme Trust, United Kingdom.
 
The crucial computing support from all WLCG partners is acknowledged gratefully, in particular from CERN, the ATLAS Tier-1 facilities at TRIUMF (Canada), NDGF (Denmark, Norway, Sweden), CC-IN2P3 (France), KIT/GridKA (Germany), INFN-CNAF (Italy), NL-T1 (Netherlands), PIC (Spain), ASGC (Taiwan), RAL (UK) and BNL (USA), the Tier-2 facilities worldwide and large non-WLCG resource providers. Major contributors of computing resources are listed in Ref.~\cite{ATL-SOFT-PUB-2021-003}.
 

\clearpage

\printbibliography
 
\clearpage 
 
\begin{flushleft}
\hypersetup{urlcolor=black}
{\Large The ATLAS Collaboration}

\bigskip

\AtlasOrcid[0000-0002-6665-4934]{G.~Aad}$^\textrm{\scriptsize 99}$,    
\AtlasOrcid[0000-0002-5888-2734]{B.~Abbott}$^\textrm{\scriptsize 125}$,    
\AtlasOrcid[0000-0002-7248-3203]{D.C.~Abbott}$^\textrm{\scriptsize 100}$,    
\AtlasOrcid[0000-0002-2788-3822]{A.~Abed~Abud}$^\textrm{\scriptsize 35}$,    
\AtlasOrcid[0000-0002-1002-1652]{K.~Abeling}$^\textrm{\scriptsize 52}$,    
\AtlasOrcid[0000-0002-2987-4006]{D.K.~Abhayasinghe}$^\textrm{\scriptsize 92}$,    
\AtlasOrcid[0000-0002-8496-9294]{S.H.~Abidi}$^\textrm{\scriptsize 28}$,    
\AtlasOrcid[0000-0002-9987-2292]{A.~Aboulhorma}$^\textrm{\scriptsize 34e}$,    
\AtlasOrcid[0000-0001-5329-6640]{H.~Abramowicz}$^\textrm{\scriptsize 158}$,    
\AtlasOrcid[0000-0002-1599-2896]{H.~Abreu}$^\textrm{\scriptsize 157}$
% ...
\bigskip
\\

% ...
$^{28}$Physics Department, Brookhaven National Laboratory, Upton NY; United States of America.\\
% ...
$^{34}$$^{(a)}$Facult\'e des Sciences Ain Chock, R\'eseau Universitaire de Physique des Hautes Energies - Universit\'e Hassan II, Casablanca;$^{(b)}$Facult\'{e} des Sciences, Universit\'{e} Ibn-Tofail, K\'{e}nitra;$^{(c)}$Facult\'e des Sciences Semlalia, Universit\'e Cadi Ayyad, LPHEA-Marrakech;$^{(d)}$LPMR, Facult\'e des Sciences, Universit\'e Mohamed Premier, Oujda;$^{(e)}$Facult\'e des sciences, Universit\'e Mohammed V, Rabat;$^{(f)}$Mohammed VI Polytechnic University, Ben Guerir; Morocco.\\
$^{35}$CERN, Geneva; Switzerland.\\
% ...
$^{52}$II. Physikalisches Institut, Georg-August-Universit\"{a}t G\"ottingen, G\"ottingen; Germany.\\
% ...
$^{92}$Department of Physics, Royal Holloway University of London, Egham; United Kingdom.\\
% ...
$^{99}$CPPM, Aix-Marseille Universit\'e, CNRS/IN2P3, Marseille; France.\\
$^{100}$Department of Physics, University of Massachusetts, Amherst MA; United States of America.\\
% ...
$^{125}$Homer L. Dodge Department of Physics and Astronomy, University of Oklahoma, Norman OK; United States of America.\\
% ...
$^{157}$Department of Physics, Technion, Israel Institute of Technology, Haifa; Israel.\\
$^{158}$Raymond and Beverly Sackler School of Physics and Astronomy, Tel Aviv University, Tel Aviv; Israel.\\
% ...
$^{a}$ Also at Borough of Manhattan Community College, City University of New York, New York NY; United States of America.\\
$^{b}$ Also at Bruno Kessler Foundation, Trento; Italy.\\
$^{c}$ Also at Center for High Energy Physics, Peking University; China.\\
$^{d}$ Also at Centro Studi e Ricerche Enrico Fermi; Italy.\\
$^{e}$ Also at CERN, Geneva; Switzerland.\\
$^{f}$ Also at D\'epartement de Physique Nucl\'eaire et Corpusculaire, Universit\'e de Gen\`eve, Gen\`eve; Switzerland.\\
% ...

\end{flushleft}


\end{document}